\documentclass[aps,pra,reprint,floatfix,showpacs,longbibliography]{revtex4-1}
\usepackage[latin9]{inputenc}
\usepackage{array}
\usepackage{multirow}
\usepackage{amsmath}
\usepackage{amssymb}

\makeatletter

\providecommand{\tabularnewline}{\\}

\usepackage{graphicx}

\newcommand*{\pr}[1]{\mathcal{#1}}
\newcommand*{\refneq}[1]{(\ref{#1})}
\newcommand*{\refeq}[1]{Eq.\ (\ref{#1})}

\makeatother

\begin{document}
\global\long\def\pr#1{\mathcal{#1}}

\title{Sum rules for spin-$1/2$ quantum gases in states with well-defined
spins: II. Spin-dependent two-body interactions.}

\author{Vladimir A. Yurovsky}

\affiliation{School of Chemistry, Tel Aviv University, 6997801 Tel Aviv, Israel}

\date{\today}
\begin{abstract}
Sums of matrix elements of spin-dependent two-body momentum-independent
interactions and sums of their products are calculated analytically
in the basis of many-body states with given total spin --- the states
built from spin and spatial wavefunctions belonging to multidimensional
irreducible representations of the symmetric group, unless the total
spin has the maximal allowed value. As in the first part of the series
{[}V. A. Yurovsky, Phys. Rev. A \textbf{91}, 053601 (2015){]}, the
sum dependence on the many-body states is given by universal factors,
which are independent of the Hamiltonians of non-interacting particles.
The sum rules are applied to perturbative analysis of energy spectra
and to calculation of two-body spin-dependent local correlations. 
\end{abstract}


\maketitle

\section*{Introduction}

The present paper continues analyses \cite{yurovsky2015} (see this
work for a more comprehensive introduction and definitions) of the
sum rules for many-body systems of indistinguishable spin-$\frac{1}{2}$
particles. The particles can be composite, e.g., atoms or molecules,
and the spin can be either a real angular momentum of the particle
or a formal spin, whose projections are attributed to the particle's
internal states (e.g. hyperfine states of atoms). In the latter case,
the spin $\frac{1}{2}$ means that only two internal states are present
in the system. This formal spin is not related to the real, physical,
spin of the particles, which can be either bosons or fermions. 

The many-body wavefunctions are represented here, as well as in \cite{yurovsky2015},
as a sum of products of the collective spin and spatial functions.
These functions depend on spin projections and coordinates, respectively,
of all particles and belong to multidimensional, non-Abelian representations
of the symmetric group (see \cite{hamermesh,elliott,kaplan,pauncz_symmetric}),
unless the total spin has the maximal allowed value. For spin-$\frac{1}{2}$
particles, the representation is unambiguously determined by the total
many-body spin. This approach is applicable to the Hamiltonians which
are separable to spin-independent and coordinate-independent parts.
It differs from the conventional approach (see \cite{landau} and
\cite{yurovsky2015}), where each particle is characterized by its
spin projection and coordinate, and the total wavefunction is symmetrized
for bosons or antisymmetrized for fermions over permutations of all
particles. The total many-body spin is undefined in this case. As
follows from the Heitler's results \cite{heitler1927} (see also \cite{yurovsky2015}),
an exact wavefunction for particles with spin-independent interactions
can not be obtained in the approach with defined individual spin projections.
Many-body states with defined total spin \cite{lieb1962,guan2009,yang2009,fang2011,daily2012,harshman2014,harshman2015,harshman2015a},
including the collective spin and spatial wavefunctions \cite{yang1967,sutherland1968,gorshkov2010,yurovsky2014},
were applied to spinor quantum gases (see \cite{myatt1997,stamper1998,ho1998,ohmi1998,pitaevskii,honerkamp2004,cazalilla2009,stamper2013,guan2013,zhang2014,scazza2014}).
Such states were also proposed for implementation of permutation quantum
computers \cite{jordan2010}. Other kings of entangled states with
non-trivial symmetry have been analyzed for quantum-degenerate gases
of spin-$1$ \cite{zhou2001} and spin-$2$ \cite{ueda2002} bosons.
A more comprehensive review is presented in Ref. \cite{yurovsky2015}. 

The well-known mean-field approach (see \cite{landau,pitaevskii}),
where the interactions between particles are replaced by the self-consistent
field, is generally applied to states with defined individual spin
projections and undefined total spins. It can be applied to certain
states with defined total spins if the interactions are spin-independent.
For bosons, the many-body mean-field wavefunction contains a single
spatial orbital, determined by the Gross-Pitaevskii equation (see
\cite{pitaevskii}). For fermions, the Hartree-Fock wavefunction (see
\cite{landau}) contains double-occupied spatial orbitals. Such wavefunctions
have the total spin $S=N/2$ or $S=0$ for $N$ spin-$\frac{1}{2}$
bosons or fermions, respectively, and describe quantum-degenerate
gases with multiple occupations of the spatial orbitals. The present
approach describes states with arbitrary total spins and is suitable
for non-degenerate gases.

For spin-independent two-body interactions between particles, sums
of matrix elements directly follow from the Heitler's results \cite{heitler1927},
while the sum of the matrix element squared moduli was calculated
in \cite{yurovsky2015}. Matrix elements of spin-dependent interactions
depend on the total spin projections of the coupled many-body states.
For spin-dependent one-body interactions with external fields, this
dependence was factorized using the Wigner-Eckart theorem and sums
of the matrix elements and their squared moduli were calculated \cite{yurovsky2015}. 

The present work is devoted to spin-dependent two-body interactions
between particles. As well as in Ref. \cite{yurovsky2015}, the matrix
elements are calculated in the basis of non-interacting particles
with single occupation of spatial modes. The interactions are expressed
in terms of irreducible spherical tensors and their matrix elements
are related to ones for the maximal allowed projections of the total
spins using the Wigner-Eckart theorem in Sec. \ref{SecSpinDep}. Sums
of these matrix elements and their squared moduli are calculated in
Sec. \ref{SecSumRules} for zero-range spin-dependent interactions.
The sum rules are applied to perturbative analysis of energy spectra
in Sec. \ref{SecEnerCorr}. Average two-body spin-dependent correlations
are analyzed in this section too. These correlations allow to distinguish
between the many-body states with defined total spins and individual
spin projections. Appendices contains calculation of sums, used in
Sec. \ref{SecSumRules}.

References to equations in the previous paper \cite{yurovsky2015}
are started here from I. The present paper, as well as \cite{yurovsky2015},
uses the following notation for the universal factors. In the ratios
of the $3j$-Wigner symbols $X_{S_{z}k}^{(S,S',q)}$, giving the dependence
on the total spin projections, $S$ and $S'$ are the maximal and
minimal total spins of the coupled states, $q$ is the rank of the
spherical tensor, and $S_{z}$ and $S_{z}+k$ are the spin projections
for $S$ and $S'$, respectively. The factors $Y_{i}^{(S,m)}[\hat{A}]$
and $Y_{i}^{(S,m)}[\hat{A},\hat{B}]$, respectively, appear in the
sum rules for the matrix elements of $\hat{A}$ and products of the
matrix elements of $\hat{A}$ and $\hat{B}$ for the maximal allowed
spin projections. Here $\hat{A}$ and $\hat{B}$ are arbitrary operators,
$S$ is the maximal total spin of the coupled states and $m$ is the
number of changed spatial quantum numbers (it is omitted if the factor
is independent of $m$). If the sum rule contains several factors
$Y$, they are specified by the subscript $i$.

\section{The spin-projection dependence\label{SecSpinDep}}

Let $|\uparrow\rangle$ and $|\downarrow\rangle$ denote two spin
states of the particles. Permutation-invariant momentum-independent
two-body interactions between particles with arbitrary spin-dependence
can be decomposed into irreducible spherical tensors of ranks $0$,
$1$, and $2$ using general relations \cite{cook1971} between irreducible
spherical and Cartesian tensors. In the present case the Cartesian
tensors are proportional to the products $\hat{s}_{\alpha}(j)\hat{s}_{\alpha'}(j')$
of the spin components of two particles, where $\alpha$ and $\alpha'$
can be either $x$, $y$, or $z$. The $z$-component of the $j$th
particle spin is 
\begin{equation}
\hat{s}_{z}(j)=\frac{1}{2}(|\uparrow(j)\rangle\langle\uparrow(j)|-|\downarrow(j)\rangle\langle\downarrow(j)|),\label{szj}
\end{equation}
 while $\hat{s}_{x}(j)=\frac{1}{2}[\hat{s}_{+}(j)+\hat{s}_{-}(j)]$
and $\hat{s}_{y}(j)=-\frac{i}{2}[\hat{s}_{+}(j)-\hat{s}_{-}(j)]$
are expressed in terms of the spin raising and lowering operators
\[
\hat{s}_{+}(j)=|\uparrow(j)\rangle\langle\downarrow(j)|,\quad\hat{s}_{-}(j)=|\downarrow(j)\rangle\langle\uparrow(j)|.
\]

There are two zero-rank tensors (spherical scalars), 
\begin{equation}
\hat{V}=\sum_{j\neq j'}V^{+}(\mathbf{r}_{j}-\mathbf{r}_{j'})\label{VindepS}
\end{equation}
 and 
\[
\hat{V}_{0}^{(0)}=-\frac{1}{\sqrt{3}}(\hat{V}_{zz}+\hat{V}_{+-}^{+}),
\]
 where 
\begin{align}
\hat{V}_{zz} & =\sum_{j\neq j'}V^{+}(\mathbf{r}_{j}-\mathbf{r}_{j'})\hat{s}_{z}(j)\hat{s}_{z}(j')\nonumber \\
\hat{V}_{+-}^{+} & =\sum_{j\neq j'}V^{+}(\mathbf{r}_{j}-\mathbf{r}_{j'})\hat{s}_{+}(j)\hat{s}_{-}(j').\label{Vpmspsm}
\end{align}
 Here 
\[
V^{\pm}(\mathbf{r})=\frac{1}{2}(V(\mathbf{r})\pm V(-\mathbf{r}))
\]
 are the even and odd parts of the two-body potential function $V(\mathbf{r})$, 

The three components of the rank $1$ spherical tensor $\hat{V}^{(1)}$
can be expressed as 
\begin{align*}
\hat{V}_{0}^{(1)} & =\frac{1}{\sqrt{2}}\sum_{j\neq j'}V^{-}(\mathbf{r}_{j}-\mathbf{r}_{j'})\hat{s}_{+}(j)\hat{s}_{-}(j')\\
\hat{V}_{\pm1}^{(1)} & =-\sum_{j\neq j'}V^{-}(\mathbf{r}_{j}-\mathbf{r}_{j'})\hat{s}_{z}(j)\hat{s}_{\pm}(j').
\end{align*}

Finally, 
\begin{align*}
\hat{V}_{0}^{(2)} & =\frac{1}{\sqrt{6}}(2\hat{V}_{zz}-\hat{V}_{+-}^{+})\\
\hat{V}_{\pm1}^{(2)} & =\mp\sum_{j\neq j'}V^{+}(\mathbf{r}_{j}-\mathbf{r}_{j'})\hat{s}_{z}(j)\hat{s}_{\pm}(j')\\
\hat{V}_{\pm2}^{(2)} & =\frac{1}{2}\sum_{j\neq j'}V^{+}(\mathbf{r}_{j}-\mathbf{r}_{j'})\hat{s}_{\pm}(j)\hat{s}_{\pm}(j')
\end{align*}
are the five components of the rank $2$ spherical tensor $\hat{V}^{(2)}$.
The even or odd rank tensors depend, respectively, on the even or
odd components $V^{\pm}(\mathbf{r})$ of the potential. Therefore
the rank 1 tensor vanish in the case of even two-body interaction,
while the scalars and rank 2 tensor vanish for the odd interaction. 

\begin{table*}
\caption{Coefficients $X_{S_{z}k}^{(S,S',2)}$ in  \refeq{VtensWE}}

\begin{tabular}{|c|c|c|c|}
\hline 
\multirow{2}{*}{$k$} & \multicolumn{3}{c|}{$S-S'$}\tabularnewline
\cline{2-4} 
 & $0$ & $1$ & $2$\tabularnewline
\hline 
0 & $\frac{3S_{z}^{2}-S(S+1)}{S(2S-1)}$ & $-\frac{S_{z}}{S-1}\sqrt{3\frac{S^{2}-S_{z}^{2}}{S(2S-1)}}$ & $\sqrt{\frac{3[(S-1)^{2}-S_{z}^{2}](S^{2}-S_{z}^{2})}{2S(S-1)(2S-1)(2S-3)}}$\tabularnewline
\hline 
1 & $-\frac{(1+2S_{z})\sqrt{6(S+S_{z}+1)(S-S_{z})}}{2S(2S-1)}$ & $\frac{S+2S_{z}+1}{S-1}\sqrt{\frac{(S-S_{z}-1)(S-S_{z})}{2S(2S-1)}}$ & $-\sqrt{\frac{(S-S_{z}-2)(S-S_{z}-1)(S^{2}-S_{z}^{2})}{S(S-1)(2S-1)(2S-3)}}$\tabularnewline
\hline 
2 & $\frac{\sqrt{6(S-S_{z}-1)(S-S_{z})(S+S_{z}+1)(S+S_{z}+2)}}{2S(2S-1)}$ & $-\frac{1}{S-1}\sqrt{\frac{[S^{2}-(S_{z}+1)^{2}](S-S_{z}-2)(S-S_{z})}{2S(2S-1)}}$ & $\sqrt{\frac{(S-S_{z}-3)(S-S_{z}-2)(S-S_{z}-1)(S-S_{z})}{4S(S-1)(2S-1)(2S-3)}}$\tabularnewline
\hline 
\end{tabular}\label{TabVtens}
\end{table*}

The interactions conserving the $z$-projection of the total many-body
spin are expressed in terms of the scalars and zero-components of
the tensors,\begin{subequations}\label{TwoBodySpinConserv} 
\begin{align}
\hat{V}_{\upuparrows} & \equiv\sum_{j\neq j'}V(\mathbf{r}_{j}-\mathbf{r}_{j'})|\uparrow(j)\rangle|\uparrow(j')\rangle\langle\uparrow(j)|\langle\uparrow(j')|\nonumber \\
 & =\sqrt{\frac{2}{3}}\hat{V}_{0}^{(2)}-\frac{1}{\sqrt{3}}\hat{V}_{0}^{(0)}+\hat{V}_{0}^{+}+\frac{1}{4}\hat{V}\label{Vupup}\\
\hat{V}_{\downdownarrows} & \equiv\sum_{j\neq j'}V(\mathbf{r}_{j}-\mathbf{r}_{j'})|\downarrow(j)\rangle|\downarrow(j')\rangle\langle\downarrow(j)|\langle\downarrow(j')|\nonumber \\
 & =\sqrt{\frac{2}{3}}\hat{V}_{0}^{(2)}-\frac{1}{\sqrt{3}}\hat{V}_{0}^{(0)}-\hat{V}_{0}^{+}+\frac{1}{4}\hat{V}\label{Vdowndown}\\
\hat{V}_{\uparrow\downarrow} & \equiv\sum_{j\neq j'}V(\mathbf{r}_{j}-\mathbf{r}_{j'})|\uparrow(j)\rangle|\downarrow(j')\rangle\langle\uparrow(j)|\langle\downarrow(j')|\nonumber \\
 & =-\sqrt{\frac{2}{3}}\hat{V}_{0}^{(2)}+\frac{1}{\sqrt{3}}\hat{V}_{0}^{(0)}-\hat{V}_{0}^{-}+\frac{1}{4}\hat{V}\label{Vupdown}\\
\hat{V}_{+-} & \equiv\sum_{j\neq j'}V(\mathbf{r}_{j}-\mathbf{r}_{j'})|\uparrow(j)\rangle|\downarrow(j')\rangle\langle\downarrow(j)|\langle\uparrow(j')|\nonumber \\
 & =-\sqrt{\frac{2}{3}}\hat{V}_{0}^{(2)}-\frac{2}{\sqrt{3}}\hat{V}_{0}^{(0)}+\sqrt{2}\hat{V}_{0}^{(1)}.\label{Vplusminus}
\end{align}
\end{subequations} Here 
\[
\hat{V}_{0}^{\pm}=\sum_{j\neq j'}V^{\pm}(\mathbf{r}_{j}-\mathbf{r}_{j'})\hat{s}_{z}(j).
\]
 are zero-components of spherical vectors (see \cite{cook1971}).
Equations \refneq{Vupup} and \refneq{Vdowndown} lead to the relation
\begin{equation}
\hat{V}_{0}^{+}=\frac{1}{2}\left(\hat{V}_{\upuparrows}-\hat{V}_{\downdownarrows}\right).\label{Vpzupdown}
\end{equation}

The interactions changing spin of one of the colliding particles,
\begin{equation}
\begin{split}\hat{V}_{-\uparrow} & \equiv\sum_{j\neq j'}V(\mathbf{r}_{j}-\mathbf{r}_{j'})|\downarrow(j)\rangle|\uparrow(j')\rangle\langle\uparrow(j)|\langle\uparrow(j')|\\
 & =\hat{V}_{-1}^{(2)}+\hat{V}_{-1}^{(1)}+\frac{1}{\sqrt{2}}\hat{V}_{-1}\\
\hat{V}_{-\downarrow} & \equiv\sum_{j\neq j'}V(\mathbf{r}_{j}-\mathbf{r}_{j'})|\downarrow(j)\rangle|\downarrow(j')\rangle\langle\uparrow(j)|\langle\downarrow(j')|\\
 & =-\hat{V}_{-1}^{(2)}-\hat{V}_{-1}^{(1)}+\frac{1}{\sqrt{2}}\hat{V}_{-1},
\end{split}
\label{TwoBodyOneSpinChange}
\end{equation}
involve also 
\[
\hat{V}_{\pm1}=\mp\frac{1}{\sqrt{2}}\sum_{j\neq j'}V(\mathbf{r}_{j}-\mathbf{r}_{j'})\hat{s}_{\pm}(j),
\]
which form a spherical vector together with $\hat{V}_{0}=\hat{V}_{z}^{+}+\hat{V}_{z}^{-}$
(see \cite{cook1971}).

The interaction 
\begin{align}
\hat{V}_{--} & \equiv\sum_{j\neq j'}V(\mathbf{r}_{j}-\mathbf{r}_{j'})|\downarrow(j)\rangle|\downarrow(j')\rangle\langle\uparrow(j)|\langle\uparrow(j')|\nonumber \\
 & =2\hat{V}_{-2}^{(2)}\label{TwoBodyTwoSpinChange}
\end{align}
changes spins of both colliding particles. Other spin-changing interactions
are obtained by the Hermitian conjugation of Eqs. \refneq{TwoBodyOneSpinChange}
and \refneq{TwoBodyTwoSpinChange}, taking into account that 
\begin{equation}
\begin{aligned}(\hat{V}_{-2}^{(2)})^{\dagger}=\hat{V}_{+2}^{(2)}, & (\hat{V}_{-1}^{(2)})^{\dagger}=-\hat{V}_{+1}^{(2)}\\
(\hat{V}_{-1}^{(1)})^{\dagger}=\hat{V}_{+1}^{(1)}, & \hat{V}_{-1}^{\dagger}=-\hat{V}_{+1}.
\end{aligned}
\label{V12conj}
\end{equation}

Consider matrix elements between wavefunctions $\Psi_{nS_{z}}^{(S)}$
with the defined projection $S_{z}$ of the total spin $S$. The explicit
form of $\Psi_{nS_{z}}^{(S)}$, given by Eq. (I.16), is not used here.
The multi-index $n$ labels different functions with the same $S$
and $S_{z}$. According to the Wigner-Eckart theorem (see \cite{edmonds}),
the matrix elements of the spherical scalars are diagonal in spins
and independent of the spin projection,
\begin{align*}
\langle\Psi_{n'S'_{z}}^{(S')}|\hat{V}|\Psi_{nS_{z}}^{(S)}\rangle & =\delta_{SS'}\delta_{S_{z}S'_{z}}\langle\Psi_{n'S}^{(S)}|\hat{V}|\Psi_{nS}^{(S)}\rangle\\
\langle\Psi_{n'S'_{z}}^{(S')}|\hat{V}_{0}^{(0)}|\Psi_{nS_{z}}^{(S)}\rangle & =\delta_{SS'}\delta_{S_{z}S'_{z}}\langle\Psi_{n'S}^{(S)}|\hat{V}_{0}^{(0)}|\Psi_{nS}^{(S)}\rangle.
\end{align*}

The matrix elements of the spherical vectors and the rank 1 tensor
follow the same relations (I.23) as the one-body interactions, 
\begin{multline}
\langle\Psi_{n'S'_{z}}^{(S')}|\hat{A}_{k}|\Psi_{nS_{z}}^{(S)}\rangle=\delta_{S'_{z}S_{z}+k}X_{S_{z}k}^{(S,S',1)}\langle\Psi_{n'S'}^{(S')}|\hat{A}_{S'-S}|\Psi_{nS}^{(S)}\rangle,\label{AvectWE}
\end{multline}
where the factors
\begin{multline*}
X_{S_{z}k}^{(S,S',q)}=(-1)^{S'-S_{z}-k}\left(\begin{array}{ccc}
S & S' & q\\
S_{z} & -S_{z}-k & k
\end{array}\right)\\
\times\left(\begin{array}{ccc}
S & S' & q\\
S & -S' & S'-S
\end{array}\right)^{-1}
\end{multline*}
are expressed in terms of the $3j$-Wigner symbols. Here $\hat{A}_{0}$
can be either $\hat{V}_{0}^{\pm}$, $\hat{V}_{0}$, or $\hat{V}_{0}^{(1)}$,
and $\hat{A}_{\pm1}$ can be $\hat{V}_{\pm1}$, or $\hat{V}_{\pm1}^{(1)}$.
According to the properties of the $3j$-Wigner symbols, the matrix
elements \refneq{AvectWE} vanish if $|S-S'|>1$ (in agreement to
the selection rules \cite{yurovsky2014}). The factors $X_{S_{z}k}^{(S,S',1)}$
for $S'\leq S$ are presented in Table I in Ref. \cite{yurovsky2015}
. Hermitian conjugate of \refeq{AvectWE}, together with \refeq{V12conj}
and relations $(\hat{V}_{0}^{\pm})^{\dagger}=\hat{V}_{0}^{\pm}$,
$(\hat{V}_{0})^{\dagger}=\hat{V}_{0}$, and $(\hat{V}_{0}^{(1)})^{\dagger}=-\hat{V}_{0}^{(1)}$,
give us the matrix elements for $S'=S+1$.

The matrix elements of the components of the rank 2 spherical tensor
$\hat{V}_{k}^{(2)}$ can be expressed in terms of the matrix elements
for the maximal allowed spin projections ($S'\leq S$) in the same
way,
\begin{multline}
\langle\Psi_{n'S'_{z}}^{(S')}|\hat{V}_{k}^{(2)}|\Psi_{nS_{z}}^{(S)}\rangle=\delta_{S'_{z}S_{z}+k}X_{S_{z}k}^{(S,S',2)}\langle\Psi_{n'S'}^{(S')}|\hat{V}_{S'-S}^{(2)}|\Psi_{nS}^{(S)}\rangle.\label{VtensWE}
\end{multline}
According to the properties of the $3j$-Wigner symbols, the matrix
elements \refneq{VtensWE} vanish if $|S-S'|>2$ (in agreement to
the selection rules \cite{yurovsky2014}). The non-vanishing factors
$X_{S_{z}k}^{(S,S',2)}$, calculated with the $3j$-Wigner symbols
\cite{landau,edmonds}, are presented in Table \ref{TabVtens}. The
symmetry properties of the $3j$-Wigner symbols \cite{landau,edmonds}
lead to the relation 
\[
X_{S_{z}-k}^{(S,S',q)}=(-1)^{S-S'+q}X_{-S_{z}k}^{(S,S',q)},
\]
providing the factors $X_{S_{z}k}^{(S,S',2)}$ for $k<0$. The matrix
elements for $S+1\leq S'\leq S+2$ are given by Hermitian conjugation
of \refeq{VtensWE}, taking into account \refeq{V12conj} and the
relation $(\hat{V}_{0}^{(2)})^{\dagger}=\hat{V}_{0}^{(2)}$ .

Thus, each permutation-invariant two-body interaction between particles
is expressed in terms of irreducible spherical tensors. Their matrix
elements for arbitrary total spin projections are related to ones
for the maximal allowed spin projections using Wigner-Eckart theorem.
The next section deals with the later matrix elements.

\section{Sum rules\label{SecSumRules}}

\subsection{Matrix elements for zero-range interactions \label{SecZeroRange}}

The sums of the matrix elements and sums of their squared moduli will
be evaluated here for zero-range spin-dependent two-body interactions
with the even potential function
\begin{equation}
V(\mathbf{r})=\delta(\mathbf{r}),\label{VzeroRan}
\end{equation}
where $\mathbf{r}$ is a $D$-dimensional vector. This function is
generally used for description of interactions of cold atoms in free
space, when $D=3$, and under tight pancake- or cigar-shape confinement,
when $D=2$ or $D=1$, respectively. In the three- and two-dimensional
cases the $\delta$-function have to be properly renormalized.

For the even potential function, we have $V^{+}(\mathbf{r})=V(\mathbf{r})$,
$V^{-}(\mathbf{r})=0$, and, therefore, $\hat{V}_{k}^{(1)}=\hat{V}_{0}^{-}=0$,
$\hat{V}_{+-}=\hat{V}_{+-}^{+}$. Besides, the identity $|\uparrow(j)\rangle\langle\uparrow(j)|+|\downarrow(j)\rangle\langle\downarrow(j)|=1$
leads to the relation 
\begin{equation}
\hat{V}_{\uparrow\downarrow}=\frac{1}{2}\left(\hat{V}-\hat{V}_{\upuparrows}-\hat{V}_{\downdownarrows}\right).\label{Vuduudd}
\end{equation}

The zero range of interaction allows to relate the matrix elements
of $\hat{V}_{\uparrow\downarrow}$ and $\hat{V}_{+-}$, defined by
\refneq{Vupdown} and \refneq{Vplusminus}, respectively, in the following
way. Wavefunctions of indistinguishable particles $\Psi$ obey to
the quantum exclusion principle $\pr P_{jj'}\Psi=\pm\Psi$ , where
the sign $+$ or $-$ is taken for bosons or fermions, respectively.
Therefore 
\begin{multline*}
\langle\Psi'|\hat{V}_{+-}|\Psi\rangle=\pm\sum_{j\neq j'}\langle\Psi'|\uparrow(j)\rangle|\downarrow(j')\rangle\delta(\mathbf{r}_{j}-\mathbf{r}_{j'})\\
\times\langle\downarrow(j)|\langle\uparrow(j')|\pr P_{jj'}|\Psi\rangle.
\end{multline*}
The permutation operator $\pr P_{jj'}$ permutes both spins and the
coordinates $\mathbf{r}_{j}$ and $\mathbf{r}_{j'}$. However, the
$\delta$-function sets $\mathbf{r}_{j}=\mathbf{r}_{j'}$ and the
coordinate permutation has no effect. Acting to the left, permutation
$\pr P_{jj'}$ of spins is restricted by the bra $\langle\downarrow(j)|\langle\uparrow(j')|$
of the interaction operator, therefore
\begin{multline*}
\langle\Psi'|\hat{V}_{+-}|\Psi\rangle=\pm\sum_{j\neq j'}\langle\Psi'|\uparrow(j)\rangle|\downarrow(j')\rangle\delta(\mathbf{r}_{j}-\mathbf{r}_{j'})\\
\times\langle\downarrow(j')|\langle\uparrow(j)|\Psi\rangle=\pm\langle\Psi'|\hat{V}_{\uparrow\downarrow}|\Psi\rangle.
\end{multline*}
Then \refneq{Vupdown} and \refneq{Vplusminus} for the even potential
function, together with \refneq{Vuduudd}, lead to
\begin{equation}
\begin{aligned}\langle\Psi'|\hat{V}_{0}^{(0)}|\Psi\rangle= & -\frac{1}{4\sqrt{3}}\langle\Psi'|\hat{V}|\Psi\rangle\\
\langle\Psi'|\hat{V}_{0}^{(2)}|\Psi\rangle= & \sqrt{\frac{3}{2}}\langle\Psi'|\frac{1}{2}\left(\hat{V}_{\upuparrows}+\hat{V}_{\downdownarrows}\right)-\frac{1}{3}\hat{V}|\Psi\rangle\\
\langle\Psi'|\hat{V}_{0}|\Psi\rangle= & \frac{1}{2}\langle\Psi'|\hat{V}_{\upuparrows}-\hat{V}_{\downdownarrows}|\Psi\rangle
\end{aligned}
\label{SphVecTens0}
\end{equation}
for bosons. Thus, matrix elements of scalars and zero-components of
the vector and tensors are expressed in term of $\hat{V}$, $\hat{V}_{\upuparrows}$,
and $\hat{V}_{\downdownarrows}$. The non-zero components of the vector
and tensor are obtained from \refneq{TwoBodyOneSpinChange} and \refneq{TwoBodyTwoSpinChange}
\begin{align}
\hat{V}_{-1}^{(2)} & =\frac{1}{2}\left(\hat{V}_{-\uparrow}-\hat{V}_{-\downarrow}\right)\nonumber \\
\hat{V}_{-1} & =\frac{1}{\sqrt{2}}\left(\hat{V}_{-\uparrow}+\hat{V}_{-\downarrow}\right)\label{SphVecTensm1}\\
\hat{V}_{-2}^{(2)} & =\frac{1}{2}\hat{V}_{--}\nonumber 
\end{align}
and their Hermitian conjugates.

For fermions, matrix elements of $\hat{V}_{\upuparrows}$, and $\hat{V}_{\downdownarrows}$
are equal to zero, in agreement with the Pauli principle --- two particles
with the same spin cannot have equal coordinates. This leads to the
zero matrix elements of the spherical vector and tensors. The matrix
elements of the spherical scalars are related as 
\begin{align}
\langle\Psi'|\hat{V}_{0}^{(0)}|\Psi\rangle & =\frac{\sqrt{3}}{4}\langle\Psi'|\hat{V}|\Psi\rangle.\label{V00Ferm}
\end{align}

The sums of the matrix elements of $\hat{V}$ and sums of their squared
moduli are derived in Ref. \cite{yurovsky2015}. Other relevant interactions
are analyzed below.

\subsection{Matrix elements for non-interacting bosons}

Let us evaluate matrix elements of $\hat{V}_{\upuparrows}$, $\hat{V}_{\downdownarrows}$,
$\hat{V}_{-\uparrow}$, $\hat{V}_{-\downarrow}$, and $\hat{V}_{--}$
between wavefunctions 
\begin{equation}
\tilde{\Psi}_{r\{n\}S_{z}}^{(S)}=f_{S}^{-1/2}\sum_{t}\tilde{\Phi}_{tr\{n\}}^{(S)}\Xi_{tS_{z}}^{(S)}\label{tilPsiSrnSz}
\end{equation}
 {[}see Eq. (I.17){]} of non-interacting bosons with defined projection
$S_{z}$ of the total spin $S$, where $f_{S}$ is the dimension of
the respective irreducible representation of the symmetric group.
The representations and functions whitin the representations, respectively,
are labeled by the standard Young tableaux $r$ and $t$ of the shape
$\lambda=[N/2+S,N/2-S]$ (see \cite{kaplan,pauncz_symmetric}). For
bosons, the spatial functions of $N$ non-interacting particles (I.11)
are represented in the form 
\begin{equation}
\tilde{\Phi}_{tr\{n\}}^{(S)}=\left(\frac{f_{S}}{N!}\right)^{1/2}\sum_{\pr P}D_{rt}^{[\lambda]}(\pr P)\prod_{j=1}^{N}\varphi_{n_{\pr Pj}}(\mathbf{r}_{j}),\label{tilPhiStrn}
\end{equation}
where $\pr P$ are permutations of $N$ symbols, $\varphi_{n}(\mathbf{r})$
are the spatial orbitals, and the relation for the Young orthogonal
matrices 
\begin{equation}
D_{tr}^{[\lambda]}(\pr P^{-1})=D_{rt}^{[\lambda]}(\pr P)\label{InvOrthMat}
\end{equation}
{[}see (I.7){]} is used (see Ref. \cite{yurovsky2015}  for other
notation). Each matrix element, e.g. the one of $\hat{V}_{\upuparrows}$,
can be decomposed into the spatial and spin parts, \begin{widetext}
\begin{equation}
\langle\tilde{\Psi}_{r'\{n'\}S'_{z}}^{(S')}|\hat{V}_{\upuparrows}|\tilde{\Psi}_{r\{n\}S_{z}}^{(S)}\rangle=(f_{S}f_{S'})^{-1/2}\sum_{t,t'}\sum_{i\neq i'}\langle\tilde{\Phi}_{t'r'\{n'\}}^{(S')}|V(\mathbf{r}_{i}-\mathbf{r}_{i'})|\tilde{\Phi}_{tr\{n\}}^{(S)}\rangle\langle\Xi_{t'S_{z}}^{(S')}|\uparrow(i)\rangle|\uparrow(i')\rangle\langle\uparrow(i)|\langle\uparrow(i')|\Xi_{tS_{z}}^{(S)}\rangle\delta_{S_{z}S'_{z}}.\label{VuutilPsi}
\end{equation}
Using \refeq{tilPhiStrn}, the spatial matrix elements can be expressed
as 
\begin{multline}
\langle\tilde{\Phi}_{t'r'\{n'\}}^{(S')}|V(\mathbf{r}_{i}-\mathbf{r}_{i'})|\tilde{\Phi}_{tr\{n\}}^{(S)}\rangle=\frac{\sqrt{f_{S}f_{S'}}}{N!}\sum_{\pr R,\pr Q}D_{r't'}^{[\lambda']}(\pr Q)D_{rt}^{[\lambda]}(\pr R)\\
\times\int d^{D}r_{i}d^{D}r_{i'}\varphi_{n'_{\pr Qi}}^{*}(\mathbf{r}_{i})\varphi_{n'_{\pr Qi'}}^{*}(\mathbf{r}_{i'})V(\mathbf{r}_{i}-\mathbf{r}_{i'})\varphi_{n_{\pr Ri}}(\mathbf{r}_{i})\varphi_{n_{\pr Ri'}}(\mathbf{r}_{i'})\prod_{i'\neq i''\neq i}\delta_{n'_{\pr Qi''},n_{\pr Ri''}}.\label{VtilPhi}
\end{multline}
 The Kronecker $\delta$-symbols appear here due to the orthogonality
of the spatial orbitals $\varphi_{n}$ and the absence of equal quantum
numbers in each of the sets $\{n\}$ and $\{n'\}$. Due to the $\delta$-symbols,
all but two spatial quantum numbers remain unchanging. Supposing that
the unchanged $n_{i''}$ are in the same positions in the sets $\{n\}$
and $\{n'\}$, one can see that the Kronecker symbols allow only $\pr Q=\pr R$
or $\pr Q=\pr R\pr P_{ii'}$. Therefore
\begin{equation}
\langle\tilde{\Phi}_{t'r'\{n'\}}^{(S')}|V(\mathbf{r}_{i}-\mathbf{r}_{i'})|\tilde{\Phi}_{tr\{n\}}^{(S)}\rangle=\frac{\sqrt{f_{S}f_{S'}}}{N!}\sum_{\pr R}D_{rt}^{[\lambda]}(\pr R)\Bigl[D_{r't'}^{[\lambda']}(\pr R)+D_{r't'}^{[\lambda']}(\pr R\pr P_{ii'})\Bigr]\langle n'_{\pr Ri'}n'_{\pr Ri}|V|n_{\pr Ri'}n_{\pr Ri}\rangle\prod_{\pr Ri'\neq j''\neq\pr Ri}\delta_{n'_{j''},n_{j''}},\label{Vspat}
\end{equation}
where for the zero-range potential \refneq{VzeroRan} the matrix elements
\[
\langle n'_{1}n'_{2}|V|n_{1}n_{2}\rangle=\langle n'_{2}n'_{1}|V|n_{1}n_{2}\rangle=\int d^{D}r\varphi_{n'_{1}}^{*}(\mathbf{r})\varphi_{n'_{2}}^{*}(\mathbf{r})\varphi_{n_{1}}(\mathbf{r})\varphi_{n_{2}}(\mathbf{r})
\]
 are invariant over permutations of $n'_{1}$ and $n'_{2}$, as well
as of $n_{1}$ and $n_{2}$.

All matrix elements are related by the Wigner-Eckart theorem to ones
for the maximal allowed spin projection, $S'_{z}=S'$, $S_{z}=S.$
The spinor matrix elements include projections of the spin wavefunctions
(I.14), derived in \cite{yurovsky2013}, 
\[
\Xi_{tS_{z}}^{(S)}=C_{SS_{z}}\sum_{\pr P}D_{t[0]}^{[\lambda]}(\pr P)\prod_{j=1}^{N/2+S_{z}}|\uparrow(\pr Pj)\rangle\prod_{j=N/2+S_{z}+1}^{N}|\downarrow(\pr Pj)\rangle,
\]
with the normalization factor 
\begin{equation}
C_{SS_{z}}=\frac{1}{(N/2+S_{z})!(N/2-S)!}\sqrt{\frac{(2S+1)(S+S_{z})!}{(N/2+S+1)(2S)!(S-S_{z})!}}.\label{CSSz}
\end{equation}
The projection

\[
\langle\uparrow(i)|\langle\uparrow(i')|\Xi_{tS}^{(S)}\rangle=C_{SS}\sum_{\pr P}D_{t[0]}^{[\lambda]}(\pr P)\sum_{l\neq l'}^{\lambda_{1}}\delta_{i,\pr Pl}\delta_{i',\pr Pl'}\prod_{l\neq j\neq l'}^{\lambda_{1}}|\uparrow(\pr Pj)\rangle\prod_{j=\lambda_{1}+1}^{N}|\downarrow(\pr Pj)\rangle
\]
(recall, $\lambda_{1,2}=N/2\pm S$) can be transformed, using substitution
$\pr P=\pr Q\pr P_{l\lambda_{1}-1}\pr P_{l'\lambda_{1}}$, to the
form
\[
\langle\uparrow(i)|\langle\uparrow(i')|\Xi_{tS}^{(S)}\rangle=C_{SS}\sum_{\pr Q}\sum_{l\neq l'}^{\lambda_{1}}D_{t[0]}^{[\lambda]}(\pr Q\pr P_{l\lambda_{1}-1}\pr P_{l'\lambda_{1}})\delta_{i,\pr Q(\lambda_{1}-1)}\delta_{i',\pr Q\lambda_{1}}\prod_{j=1}^{\lambda_{1}-2}|\uparrow(\pr Qj)\rangle\prod_{j=\lambda_{1}+1}^{N}|\downarrow(\pr Qj)\rangle.
\]
The permutations $\pr P_{l\lambda_{1}-1}$ and $\pr P_{l'\lambda_{1}}$
permute symbols in the first row of the Young tableau $[0]$. Therefore,
$D_{t[0]}^{[\lambda]}(\pr Q\pr P_{l\lambda_{1}-1}\pr P_{l'\lambda_{1}})=D_{t[0]}^{[\lambda]}(\pr Q)$
{[}see Eq. (I.8){]} , the summand in the equation above is independent
of $l$ and $l'$, and the projection can be expressed as
\[
\langle\uparrow(i)|\langle\uparrow(i')|\Xi_{tS}^{(S)}\rangle=\lambda_{1}(\lambda_{1}-1)C_{SS}\sum_{\pr Q}D_{t[0]}^{[\lambda]}(\pr Q)\delta_{i,\pr Q(\lambda_{1}-1)}\delta_{i',\pr Q\lambda_{1}}\prod_{j=1}^{\lambda_{1}-2}|\uparrow(\pr Qj)\rangle\prod_{j=\lambda_{1}+1}^{N}|\downarrow(\pr Qj)\rangle.
\]
The projections involved into matrix elements for other interactions
are evaluated in the same way,
\begin{align*}
\langle\downarrow(i)|\langle\downarrow(i')|\Xi_{tS}^{(S)}\rangle & =\lambda_{2}(\lambda_{2}-1)C_{SS}\sum_{\pr Q}D_{t[0]}^{[\lambda]}(\pr Q)\delta_{i,\pr Q(\lambda_{1}+1)}\delta_{i',\pr Q(\lambda_{1}+2)}\prod_{j=1}^{\lambda_{1}}|\uparrow(\pr Qj)\rangle\prod_{j=\lambda_{1}+3}^{N}|\downarrow(\pr Qj)\rangle\\
\langle\uparrow(i)|\langle\downarrow(i')|\Xi_{tS}^{(S)}\rangle & =\lambda_{1}\lambda_{2}C_{SS}\sum_{\pr Q}D_{t[0]}^{[\lambda]}(\pr Q)\delta_{i,\pr Q\lambda_{1}}\delta_{i',\pr Q(\lambda_{1}+1)}\prod_{j=1}^{\lambda_{1}-1}|\uparrow(\pr Qj)\rangle\prod_{j=\lambda_{1}+2}^{N}|\downarrow(\pr Qj)\rangle.
\end{align*}

In the spin matrix elements of $\hat{V}_{\upuparrows}$,
\begin{multline*}
\langle\Xi_{t'S}^{(S)}|\uparrow(i)\rangle|\uparrow(i')\rangle\langle\uparrow(i)|\langle\uparrow(i')|\Xi_{tS}^{(S)}\rangle=\left[\lambda_{1}(\lambda_{1}-1)C_{SS}\right]^{2}\sum_{\pr Q}D_{t[0]}^{[\lambda]}(\pr Q)\delta_{i,\pr Q(\lambda_{1}-1)}\delta_{i',\pr Q\lambda_{1}}\sum_{\pr R}D_{t'[0]}^{[\lambda]}(\pr R)\delta_{i,\pr R(\lambda_{1}-1)}\delta_{i',\pr R\lambda_{1}}\\
\times\sum_{\pr P',\pr P''}\delta_{\pr R,\pr Q\pr P'\pr P''},
\end{multline*}
the last Kronecker symbol appears due to orthogonality of the spin
states and means that the permutations $\pr R$ and $\pr Q$ can be
different by permutations of particles in the same spin state. They
are the permutations $\pr P'$ of the first $\lambda_{1}-2$ symbols
and $\pr P''$ of the last $\lambda_{2}$ ones. As the permutations
$\pr P'$ and $\pr P''$ do not permute symbols between rows in the
Young tableau $[0]$, the equality $D_{t'[0]}^{[\lambda]}(\pr Q\pr P'\pr P'')=D_{t'[0]}^{[\lambda]}(\pr Q)$
{[}see Eq. (I.8){]} can be applied, leading to
\[
\langle\Xi_{t'S}^{(S)}|\uparrow(i)\rangle|\uparrow(i')\rangle\langle\uparrow(i)|\langle\uparrow(i')|\Xi_{tS}^{(S)}\rangle=\lambda_{1}!\lambda_{2}!\lambda_{1}(\lambda_{1}-1)C_{SS}^{2}\sum_{\pr Q}D_{t[0]}^{[\lambda]}(\pr Q)D_{t'[0]}^{[\lambda]}(\pr Q)\delta_{i,\pr Q(\lambda_{1}-1)}\delta_{i',\pr Q\lambda_{1}}.
\]
Let us substitute this equation and \refeq{Vspat} into \refeq{VuutilPsi},
perform the summation over $t$ and $t'$, using the relation 
\begin{equation}
\sum_{t}D_{r't}^{[\lambda]}(\pr P)D_{tr}^{[\lambda]}(\pr Q)=D_{r'r}^{[\lambda]}(\pr P\pr Q),\label{RepProd}
\end{equation}
{[}see Eq. (I.6){]} and substitute $\pr P=\pr Q^{-1}\pr R^{-1}$,
$j=\pr Ri$, and $j'=\pr Ri'$. Then the Kronecker symbols lead to
$\pr Pj=\pr Q^{-1}i=\lambda_{1}-1$ and $\pr Pj'=\pr Q^{-1}i'=\lambda_{1}$.
Equations 
\begin{equation}
\pr P\pr P_{ii'}\pr P^{-1}=\pr P_{\pr Pi\pr Pi'}\label{TransfTransp}
\end{equation}
 (see \cite{pauncz_symmetric}) and (I.8) lead then to $D_{r'[0]}^{[\lambda]}(\pr R\pr P_{ii'}Q)=D_{r'[0]}^{[\lambda]}(\pr P^{-1}\pr Q^{-1}\pr P_{ii'}Q)=D_{r'[0]}^{[\lambda]}(\pr P^{-1}\pr P_{\lambda_{1}\lambda_{1}-1})=D_{r'[0]}^{[\lambda]}(\pr P^{-1})$.
Then using \refeq{InvOrthMat} we get
\begin{equation}
\langle\tilde{\Psi}_{r'\{n'\}S'}^{(S')}|\hat{V}_{\upuparrows}|\tilde{\Psi}_{r\{n\}S}^{(S)}\rangle=2\delta_{S'S}\lambda_{1}!\lambda_{2}!\lambda_{1}(\lambda_{1}-1)C_{SS}^{2}V_{r'\{n'\}r\{n\}}^{[\lambda][\lambda]}(\lambda_{1}-1,\lambda_{1})\label{Vuunpn}
\end{equation}
with
\begin{equation}
V_{r'\{n'\}r\{n\}}^{[\lambda'][\lambda]}(l,l')=\sum_{j\neq j'}\sum_{\pr P}D_{[0]r'}^{[\lambda']}(\pr P)D_{[0]r}^{[\lambda]}(\pr P)\delta_{l,\pr Pj}\delta_{l',\pr Pj'}\langle n'_{j}n'_{j'}|V|n_{j}n_{j'}\rangle\prod_{j'\neq j''\neq j}\delta_{n'_{j''},n_{j''}}.\label{Vlplrpnprn}
\end{equation}

Matrix elements of other operators are calculated in the same way,
\begin{equation}
\begin{aligned}\langle\tilde{\Psi}_{r'\{n'\}S'}^{(S')}|\hat{V}_{\downdownarrows}|\tilde{\Psi}_{r\{n\}S}^{(S)}\rangle & =2\delta_{S'S}\lambda_{1}!\lambda_{2}!\lambda_{2}(\lambda_{2}-1)C_{SS}^{2}V_{r'\{n'\}r\{n\}}^{[\lambda][\lambda]}(\lambda_{1}+1,\lambda_{1}+2)\\
\langle\tilde{\Psi}_{r'\{n'\}S'}^{(S')}|\hat{V}_{-\uparrow}|\tilde{\Psi}_{r\{n\}S}^{(S)}\rangle & =2\delta_{S'S-1}\lambda_{1}!\lambda_{2}!(\lambda_{1}-1)(\lambda_{2}+1)C_{SS}C_{S-1S-1}V_{r'\{n'\}r\{n\}}^{[\lambda'][\lambda]}(\lambda_{1}-1,\lambda_{1})\\
\langle\tilde{\Psi}_{r'\{n'\}S'}^{(S')}|\hat{V}_{-\downarrow}|\tilde{\Psi}_{r\{n\}S}^{(S)}\rangle & =2\delta_{S'S-1}\lambda_{1}!\lambda_{2}!\lambda_{2}(\lambda_{2}+1)C_{SS}C_{S-1S-1}V_{r'\{n'\}r\{n\}}^{[\lambda'][\lambda]}(\lambda_{1},\lambda_{1}+1)\\
\langle\tilde{\Psi}_{r'\{n'\}S'}^{(S')}|\hat{V}_{--}|\tilde{\Psi}_{r\{n\}S}^{(S)}\rangle & =2\delta_{S'S-2}\lambda_{1}!\lambda_{2}!(\lambda_{2}+1)(\lambda_{2}+2)C_{SS}C_{S-2S-2}V_{r'\{n'\}r\{n\}}^{[\lambda'][\lambda]}(\lambda_{1}-1,\lambda_{1}),
\end{aligned}
\label{Vmatel}
\end{equation}
\end{widetext}where $\lambda'=[N/2+S',N/2-S']$.

\subsection{Sums of the matrix elements and their squares and products}

For the matrix elements which are diagonal in the total spin and $r$,
one can calculate their sums and write out them in the form\begin{subequations}\label{Y1S2}
\begin{multline}
\sum_{r}\langle\tilde{\Psi}_{r\{n'\}S}^{(S)}|\hat{V}_{a}|\tilde{\Psi}_{r\{n\}S}^{(S)}\rangle=Y^{(S)}[\hat{V}_{a}]\frac{2f_{S}}{N(N-1)}\\
\times\sum_{j<j'}\langle n'_{j}n'_{j'}|V|n_{j}n_{j'}\rangle\prod_{j'\neq j''\neq j}\delta_{n'_{j''},n_{j''}},\label{SumVSS}
\end{multline}
where $\hat{V}_{a}$ is any two-body interaction, which does not change
the spin projection. In each term of the sum over $j$ and $j'$,
only two spatial quantum numbers can be changed. For the operators
$\hat{V}_{\upuparrows}$ and $\hat{V}_{\downdownarrows}$ the factors
\[
Y^{(S)}[\hat{V}_{\upuparrows}]=2\lambda_{1}(\lambda_{1}-1),\quad Y^{(S)}[\hat{V}_{\downdownarrows}]=2\lambda_{2}(\lambda_{2}-1),
\]
calculated using the Eqs. \refneq{Vuunpn}, \refneq{Vmatel}, \refneq{CSSz},
\refneq{Vlplrpnprn}, \refneq{RepProd}, and \refneq{InvOrthMat},
are proportional to the numbers $\lambda_{1}(\lambda_{1}-1)$ and
$\lambda_{2}(\lambda_{2}-1)$ of particle pairs with spins $\uparrow$
and $\downarrow$, respectively. For the spherical tensor components,
the factors are calculated with \refeq{SphVecTens0}, 
\begin{equation}
Y^{(S)}[\hat{V}_{0}^{(2)}]=\sqrt{\frac{2}{3}}S(2S-1),\quad Y^{(S)}[\hat{V}_{0}]=2S(N-1).
\end{equation}
The sum of the matrix elements of the spin-independent interactions
(I.45a) can be expressed for bosons and the zero-range potentials
\refneq{VzeroRan} in the form \refneq{SumVSS} too with
\begin{equation}
Y^{(S)}[\hat{V}]=\frac{3}{2}N(N-2)+2S(S+1).
\end{equation}
\end{subequations}

The sums of squared moduli of the matrix elements \refneq{Vuunpn}
and \refneq{Vmatel} and their products are propotional to the sums
of products of the functions \refneq{Vlplrpnprn}, which can be expressed
as\begin{widetext} 
\begin{multline}
\sum_{r,r'}V_{r'\{n'\}r\{n\}}^{[\lambda'][\lambda]}(l_{1},l_{1}')V_{r'\{n\}r\{n'\}}^{[\lambda'][\lambda]}(l_{2},l_{2}')=\sum_{j_{1}\neq j_{1}'}\sum_{j_{2}\neq j_{2}'}\varSigma_{j_{1}j'_{1}j_{2}j'_{2}}^{(S',S)}(l_{1},l'_{1},l_{2},l'_{2})\\
\times\langle n'_{j_{1}}n'_{j_{1}'}|V|n_{j_{1}}n_{j_{1}'}\rangle\prod_{j_{1}'\neq j_{1}''\neq j_{1}}\delta_{n'_{j_{1}''},n_{j_{1}''}}\langle n'_{j_{2}}n'_{j_{2}'}|V|n_{j_{2}}n_{j_{2}'}\rangle^{*}\prod_{j_{2}'\neq j_{2}''\neq j_{2}}\delta_{n'_{j_{2}''},n_{j_{2}''}},\label{SumVuu2}
\end{multline}
where
\begin{equation}
\varSigma_{j_{1}j'_{1}j_{2}j'_{2}}^{(S',S)}(l_{1},l'_{1},l_{2},l'_{2})=\sum_{r,r'}\sum_{\pr P}D_{[0]r'}^{[\lambda']}(\pr P)D_{[0]r}^{[\lambda]}(\pr P)\delta_{l_{1},\pr Pj_{1}}\delta_{l'_{1},\pr Pj_{1}'}\sum_{\pr Q}D_{[0]r'}^{[\lambda']}(\pr Q)D_{[0]r}^{[\lambda]}(\pr Q)\delta_{l_{2},\pr{\pr Q}j_{2}}\delta_{l'_{2},\pr{\pr Q}j_{2}'}\label{SigmaSSp4}
\end{equation}

The sums of squared moduli contain different functions $\varSigma_{j_{1}j'_{1}j_{2}j'_{2}}^{(S',S)}$,
namely $\varSigma_{j_{1}j'_{1}j_{2}j'_{2}}^{(S,S)}(\lambda_{1}-1,\lambda_{1},\lambda_{1}-1,\lambda_{1})$
for $\hat{V}_{\upuparrows}$, $\varSigma_{j_{1}j'_{1}j_{2}j'_{2}}^{(S,S)}(\lambda_{1}+1,\lambda_{1}+2,\lambda_{1}+1,\lambda_{1}+2)$
for $\hat{V}_{\downdownarrows}$, $\varSigma_{j_{1}j'_{1}j_{2}j'_{2}}^{(S-1,S)}(\lambda_{1}-1,\lambda_{1},\lambda_{1}-1,\lambda_{1})$
for $\hat{V}_{-\uparrow}$, $\varSigma_{j_{1}j'_{1}j_{2}j'_{2}}^{(S-1,S)}(\lambda_{1},\lambda_{1}+1,\lambda_{1},\lambda_{1}+1)$
for $\hat{V}_{-\downarrow}$, and $\varSigma_{j_{1}j'_{1}j_{2}j'_{2}}^{(S-2,S)}(\lambda_{1}-1,\lambda_{1},\lambda_{1}-1,\lambda_{1})$
for $\hat{V}_{--}$. Calculation of the sums of squared moduli of
the matrix elements of spherical vectors and tensors with \refneq{SphVecTens0}
and \refneq{SphVecTensm1} requires also sums over $r$ and $r'$
of the products of matrix elements. The latter sums contain $\varSigma_{j_{1}j'_{1}j_{2}j'_{2}}^{(S,S)}(\lambda_{1}-1,\lambda_{1},\lambda_{1}+1,\lambda_{1}+2)$
for products of the matrix elements of $\hat{V}_{\upuparrows}$ by
$\hat{V}_{\downdownarrows}$ and $\varSigma_{j_{1}j'_{1}j_{2}j'_{2}}^{(S-1,S)}(\lambda_{1}-1,\lambda_{1},\lambda_{1},\lambda_{1}+1)$
for $\hat{V}_{-\uparrow}$ by $\hat{V}_{-\downarrow}$. The sums $\varSigma_{j_{1}j'_{1}j_{2}j'_{2}}^{(S',S)}$
are calculated in Appendix \ref{AppSigmaSSp}. 

The sums of products of matrix elements of $\hat{V}$ {[}expressed
by (I.44) with the zero-range potential function \refneq{VzeroRan}{]}
by $\hat{V}_{\upuparrows}$ or $\hat{V}_{\downdownarrows}$ are proportional
to the sum 
\begin{multline}
\sum_{r,r'}V_{r'\{n'\}r\{n\}}^{[\lambda][\lambda]}(l,l')\langle\tilde{\Psi}_{r\{n\}S}^{(S)}|\hat{V}|\tilde{\Psi}_{r'\{n'\}S}^{(S)}\rangle=\sum_{j_{1}\neq j_{1}'}\sum_{j_{2}\neq j_{2}'}\left[(N-2)!+\varSigma_{j_{1}j'_{1}j_{2}j'_{2}}^{(S)}(l,l')\right]\\
\times\langle n'_{j_{1}}n'_{j_{1}'}|V|n_{j_{1}}n_{j_{1}'}\rangle\prod_{j_{1}'\neq j_{1}''\neq j_{1}}\delta_{n'_{j_{1}''},n_{j_{1}''}}\langle n'_{j_{2}}n'_{j_{2}'}|V|n_{j_{2}}n_{j_{2}'}\rangle^{*}\prod_{j_{2}'\neq j_{2}''\neq j_{2}}\delta_{n'_{j_{2}''},n_{j_{2}''}},\label{SumVuuV}
\end{multline}
\end{widetext}where
\begin{multline}
\varSigma_{j_{1}j'_{1}j_{2}j'_{2}}^{(S)}(l,l')=\sum_{r,r'}\sum_{\pr P}D_{[0]r'}^{[\lambda]}(\pr P)D_{[0]r}^{[\lambda]}(\pr P)\delta_{l,\pr Pj_{1}}\delta_{l',\pr Pj_{1}'}\\
\times D_{rr'}^{[\lambda]}(\pr P_{j_{2}j'_{2}})\label{SigmaS4}
\end{multline}
is calculated in Appendix \ref{AppS4}. Equation \refneq{SumVuuV}
contains $\varSigma_{j_{1}j'_{1}j_{2}j'_{2}}^{(S)}(\lambda_{1}-1,\lambda_{1})$
and $\varSigma_{j_{1}j'_{1}j_{2}j'_{2}}^{(S)}(\lambda_{1}+1,\lambda_{1}+2)$
for $\hat{V}_{\upuparrows}$ and $\hat{V}_{\downdownarrows}$, respectively.
The sums of squared moduli of the matrix elements and their products
are expressed in different forms if the set of spatial quantum numbers
is changed ($\{n\}\neq\{n'\}$) or conserved ($\{n\}=\{n'\}$).

\subsection{Changing set of spatial quantum numbers}

If the sets of spatial quantum numbers $\{n\}$ and $\{n'\}$ are
different by two elements,, the product of Kronecker symbols in \refneq{SumVuu2}
and \refneq{SumVuuV} does not vanish only if either $j_{1}=j_{2}$,
$j'_{1}=j'_{2}$ or $j_{1}=j'_{2}$, $j'_{1}=j_{2}$ . Since $\varSigma_{jj'jj'}^{(S',S)}(l_{1},l'_{1},l_{2},l'_{2})$
is independent of particular values of $j$ and $j'$ (see Appendix
\ref{AppSigmaSSp}), the sum \refneq{SumVuu2} attains the form
\begin{multline*}
\sum_{r,r'}V_{r'\{n'\}r\{n\}}^{[\lambda'][\lambda]}(l_{1},l_{1}')V_{r'\{n\}r\{n'\}}^{[\lambda'][\lambda]}(l_{2},l_{2}')\\
=2\varSigma_{2}^{(S',S)}(l_{1},l'_{1},l_{2},l'_{2})\sum_{j<j'}|\langle n'_{j}n'_{j'}|V|n_{j}n_{j'}\rangle|^{2}\prod_{j'\neq j''\neq j}\delta_{n'_{j''},n_{j''}}
\end{multline*}
with
\begin{equation}
\varSigma_{2}^{(S',S)}(l_{1},l'_{1},l_{2},l'_{2})=\varSigma_{jj'jj'}^{(S',S)}(l_{1},l'_{1},l_{2},l'_{2})+\varSigma_{jj'jj'}^{(S',S)}(l_{1},l'_{1},l'_{2},l_{2}).\label{SigmaSSp2}
\end{equation}

Then for any two-body spin-dependent interactions $\hat{V}_{a}$ and
$\hat{V}_{b}$, the sums of squared moduli of the matrix elements
and their products can be written out in the form\begin{subequations}\label{Y2S2}
\begin{multline}
\sum_{r,r'}\langle\tilde{\Psi}_{r'\{n'\}S'}^{(S')}|\hat{V}_{a}|\tilde{\Psi}_{r\{n\}S}^{(S)}\rangle\langle\tilde{\Psi}_{r'\{n'\}S'}^{(S')}|\hat{V}_{b}|\tilde{\Psi}_{r\{n\}S}^{(S)}\rangle^{*}\\
=Y^{(S,2)}[\hat{V}_{a},\hat{V}_{b}]\frac{2f_{S'}}{N(N-1)}\sum_{j<j'}|\langle n'_{j}n'_{j'}|V|n_{j}n_{j'}\rangle|^{2}\\
\times\prod_{j'\neq j''\neq j}\delta_{n'_{j''},n_{j''}}\label{SumVaVbnpn}
\end{multline}
with $S'\leq S$. Each term in the sum above changes two of the spatial
quantum numbers, conserving other ones. Since $S$ and $S'$ are equal
to the spin projections, $S'$ is unambiguously determined by the
operators $\hat{V}_{a}$ and $\hat{V}_{b}$, such that $S'=S$ for
$\hat{V}$, $\hat{V}_{\upuparrows}$, and $\hat{V}_{\downdownarrows}$
and $S'=S+k$ for $\hat{V}_{k}^{(2)}$ and $\hat{V}_{k}$. The factors
$Y^{(S,2)}[\hat{V}_{a},\hat{V}_{b}]$ are expressed in terms of the
sums $\varSigma_{2}^{(S',S)}(l_{1},l'_{1},l_{2},l'_{2})$. For example,
the factor $Y^{(S,2)}[\hat{V}_{\upuparrows},\hat{V}_{\upuparrows}]$
takes the form
\begin{multline*}
Y^{(S,2)}[\hat{V}_{\upuparrows},\hat{V}_{\upuparrows}]=4\left[\lambda!\lambda_{2}!\lambda_{1}(\lambda_{1}-1)C_{SS}^{2}\right]^{2}\frac{N(N-1)}{f_{S}}\\
\times\varSigma_{2}^{(S,S)}(\lambda_{1}-1,\lambda_{1},\lambda_{1}-1,\lambda_{1}),
\end{multline*}

Similarly, the sum \refneq{SumVuuV} leads to the factor
\begin{multline*}
Y^{(S,2)}[\hat{V}_{\upuparrows},\hat{V}]=4\lambda_{1}!\lambda_{2}!\lambda_{1}(\lambda_{1}-1)C_{SS}^{2}\frac{N(N-1)}{f_{S}}\\
\times\left[(N-2)!+\varSigma_{jj'jj'}^{(S)}(\lambda_{1}-1,\lambda_{1})\right].
\end{multline*}

The factors in the sums of squared moduli and products of matrix elements
of other operators are expressed in a similar form. Explicit expressions
for the factors are obtained using the normalization factors \refneq{CSSz}
and sums $\varSigma_{jj'jj'}^{(S',S)}$ and $\varSigma_{jj'jj'}^{(S)}$,
calculated in Appendices \ref{AppSigmaSSp} and \ref{AppS4}, respectively.
For example,
\begin{multline*}
Y^{(S,2)}[\hat{V}_{\upuparrows},\hat{V}_{\upuparrows}]=\frac{2}{2S+3}\biggl[N^{2}(2S+1)\\
+2\frac{N(4S^{3}+8S^{2}+S-1)+2S(2S^{3}+5S^{2}-5)}{S+1}\biggr].
\end{multline*}
Equations \refneq{SphVecTens0} and \refneq{SphVecTensm1} lead to
the following factors in the sums of squared moduli of the matrix
elements of spherical tensor components and their products
\begin{align}
Y^{(S,2)}[\hat{V}_{0}^{(2)},\hat{V}_{0}^{(2)}] & =\frac{S(2S-1)}{6(2S+3)}\left(3\frac{(N+2)^{2}}{S+1}-4S\right)\\
Y^{(S,2)}[\hat{V}_{0},\hat{V}_{0}] & =S\left(4S+\frac{N^{2}-4}{S+1}\right)
\end{align}
 
\begin{align}
Y^{(S,2)}[\hat{V}_{0}^{(2)},\hat{V}_{0}] & =\sqrt{\frac{2}{3}}\frac{(N+2)S(2S-1)}{S+1}\\
Y^{(S,2)}[\hat{V}_{0}^{(2)},\hat{V}] & =4\sqrt{\frac{2}{3}}S(2S-1)\\
Y^{(S,2)}[\hat{V}_{0},\hat{V}] & =8S(N-1)
\end{align}
\begin{align}
Y^{(S,2)}[\hat{V}_{-1}^{(2)},\hat{V}_{-1}^{(2)}] & =\frac{(N+2)(N-2S+2)(S-1)}{2(S+1)}\\
Y^{(S,2)}[\hat{V}_{-1},\hat{V}_{-1}] & =(N-2)(N-2S+2)\\
Y^{(S,2)}[\hat{V}_{-1}^{(2)},\hat{V}_{-1}] & =\sqrt{2}(N-2S+2)(S-1)\\
Y^{(S,2)}[\hat{V}_{-2}^{(2)},\hat{V}_{-2}^{(2)}] & =\frac{1}{2}(N-2S+2)(N-2S+4)
\end{align}
The sum of squared moduli of the matrix elements of the spin-independent
interactions (I.47a) can be expressed in the case of bosons with the
zero-range potentials \refneq{VzeroRan} in the form \refneq{SumVaVbnpn}
too with
\begin{equation}
Y^{(S,2)}[\hat{V},\hat{V}]=6N(N-2)+8S(S+1).
\end{equation}
\end{subequations} The case of a single changed quantum number will
be considered elsewhere.

\subsection{Conserving set of spatial quantum numbers}

If the set of spatial quantum numbers is unchanged, $\{n\}=\{n'\}$,
the Kronecker symbols in sums \refneq{SumVuu2} and \refneq{SumVuuV}
are equal to one for any $j_{1}$, $j_{2}$, $j'_{1}$, and $j'_{2}$.
Then these sums contain\begin{widetext} 
\begin{equation}
\sum_{j_{1}\neq j_{1}'}\sum_{j_{2}\neq j_{2}'}\varSigma_{j_{1}j'_{1}j_{2}j'_{2}}V_{j_{1}j'_{1}}V_{j_{2}j'_{2}}=\varSigma_{4}(\sum_{j\neq j'}V_{jj'})^{2}+(\varSigma_{3}-4\varSigma_{4})\sum_{j'\neq j\neq j''}V_{jj'}V_{jj''}+(\varSigma_{2}-\varSigma_{3}+2\varSigma_{4})\sum_{j\neq j'}V_{jj'}^{2},\label{SumSigVV}
\end{equation}
where the matrix elements of zero-range interactions \refneq{VzeroRan}
are symmetric over permutations of $n_{j}$ and $n_{j'}$, 
\begin{equation}
V_{jj'}=\langle n_{j}n_{j'}|V|n_{j}n_{j'}\rangle=\langle n_{j'}n_{j}|V|n_{j}n_{j'}\rangle=\langle n_{j}n_{j'}|V|n_{j'}n_{j}\rangle=\int d^{D}r|\varphi_{n_{j}}(\mathbf{r})\varphi_{n_{j'}}(\mathbf{r})|^{2}\label{Vjjp}
\end{equation}
and $\varSigma_{j_{1}j'_{1}j_{2}j'_{2}}$ can be either $\varSigma_{j_{1}j'_{1}j_{2}j'_{2}}^{(S',S)}(l_{1},l'_{1},l_{2},l'_{2})$
or $\varSigma_{j_{1}j'_{1}j_{2}j'_{2}}^{(S)}(l,l')$ {[}see Eqs. \refneq{SigmaSSp4}
and \refneq{SigmaS4}{]} with arbitrary superscripts and arguments.
These functions depend on relations between their subscripts, rather
than the subscript specific values (see Appendices \ref{AppSigmaSSp}
and \ref{AppS4}). Then $\varSigma_{4}=\varSigma_{j_{1}j'_{1}j_{2}j'_{2}}$
for $j_{1}\neq j_{2}\neq j'_{1}$ and $j_{1}\neq j'_{2}\neq j'_{1}$,
$\varSigma_{3}=\varSigma_{jj'jj''}+\varSigma_{jj'j''j}+\varSigma_{j'jjj''}+\varSigma_{j'jj''j}$
for $j'\neq j\neq j''\neq j'$, and $\varSigma_{2}=\varSigma_{jj'jj'}+\varSigma_{jj'j'j}$
for $j\neq j'$. The sum \refneq{SumSigVV} can be further transformed
to
\begin{multline*}
\sum_{j_{1}\neq j_{1}'}\sum_{j_{2}\neq j_{2}'}\varSigma_{j_{1}j'_{1}j_{2}j'_{2}}V_{j_{1}j'_{1}}V_{j_{2}j'_{2}}=N(N-1)[(N-2)(N-3)\varSigma_{4}+(N-2)\varSigma_{3}+\varSigma_{2}]\langle V\rangle^{2}+N(N-1)^{2}(\varSigma_{3}-4\varSigma_{4})\langle\Delta_{1}V\rangle^{2}\\
+N(N-1)(\varSigma_{2}-\varSigma_{3}+2\varSigma_{4})(\langle\Delta_{2}V\rangle^{2}.
\end{multline*}
\end{widetext}Here 
\begin{equation}
\langle V\rangle=\frac{2}{N(N-1)}\sum_{j<j'}V_{jj'}\label{Vaver}
\end{equation}
is the average value of the matrix elements \refneq{Vjjp} and
\begin{equation}
\begin{aligned}\langle\Delta_{1}V\rangle^{2} & =\frac{1}{N}\sum_{j=1}^{N}\left(\frac{1}{N-1}\sum_{j'\neq j}V_{jj'}-\langle V\rangle\right)^{2}\\
\langle\Delta_{2}V\rangle^{2} & =\frac{2}{N(N-1)}\sum_{j<j'}(V_{jj'}-\langle V\rangle)^{2}
\end{aligned}
\label{DeltaV}
\end{equation}
measure their average deviations {[}in consistency with (I.48){]}.
Then the sums of squared moduli of the matrix elements and their products
can be written out in the form\begin{widetext}\begin{subequations}\label{Y2SSp2nn}
\begin{equation}
\sum_{r,r'}\langle\tilde{\Psi}_{r'\{n\}S'}^{(S')}|\hat{V}_{a}|\tilde{\Psi}_{r\{n\}S}^{(S)}\rangle\langle\tilde{\Psi}_{r'\{n\}S'}^{(S')}|\hat{V}_{b}|\tilde{\Psi}_{r\{n\}S}^{(S)}\rangle^{*}=f_{S'}\left(Y_{0}^{(S,0)}[\hat{V}_{a},\hat{V}_{b}]\langle V\rangle^{2}+Y_{1}^{(S,0)}[\hat{V}_{a},\hat{V}_{b}]\langle\Delta_{1}V\rangle^{2}+Y_{2}^{(S,0)}[\hat{V}_{a},\hat{V}_{b}]\langle\Delta_{2}V\rangle^{2}\right)\label{SumVaVbnn}
\end{equation}
with $S'\leq S$. The factors $Y_{i}^{(S,0)}$ here are calculated
using results of Appendices \ref{AppSigmaSSp}, \ref{AppS4}, and
\ref{AppSSpjj}. The first factor can be represented for all considered
$\hat{V}_{a}$ and $\hat{V}_{b}$ as 
\begin{equation}
Y_{0}^{(S,0)}[\hat{V}_{a},\hat{V}_{b}]=\delta_{SS'}Y^{(S)}[\hat{V}_{a}]Y^{(S)}[\hat{V}_{b}]\label{Y20SS2nn}
\end{equation}
 {[}see \refeq{Y1S2}{]}. Since $S'-S$ is unambiguously determined
by $\hat{V}_{a}$ and $\hat{V}_{b}$ ($S'=S$ for $\hat{V}$ and $S'=S+k$
for $\hat{V}_{k}^{(2)}$ and $\hat{V}_{k}$), $Y_{0}^{(S,0)}[\hat{V}_{a},\hat{V}_{b}]=0$
for $k\neq0$ components of the spherical vectors and tensors, when
$S\neq S'$. Other factors are expressed as
\begin{align}
Y_{1}^{(S,0)}[\hat{V}_{0}^{(2)},\hat{V}_{0}^{(2)}] & =-\frac{(N-1)(N-2S)(N+2S+2)S(2S-1)}{3(N-2)(N-3)(2S+3)}\left(3\frac{N+3}{S+1}-8S\right)\\
Y_{2}^{(S,0)}[\hat{V}_{0}^{(2)},\hat{V}_{0}^{(2)}] & =\frac{(N-2S)(N+2S+2)S(2S-1)}{6(N-2)(N-3)(2S+3)}\left(3N\frac{N-1}{S+1}-8S\right)\\
Y_{1}^{(S,0)}[\hat{V}_{0},\hat{V}_{0}] & =\frac{(N-1)(N-2S)(N+2S+2)S}{S+1}
\end{align}
 
\begin{align}
Y_{1}^{(S,0)}[\hat{V}_{0}^{(2)},\hat{V}_{0}] & =\sqrt{\frac{2}{3}}\frac{(N-1)(N-2S)(N+2S+2)S(2S-1)}{(N-2)(S+1)}\\
Y_{1}^{(S,0)}[\hat{V}_{0}^{(2)},\hat{V}] & =2\sqrt{\frac{2}{3}}\frac{(N-1)(N-2S)(N+2S+2)S(2S-1)}{(N-2)(N-3)}\\
Y_{2}^{(S,0)}[\hat{V}_{0}^{(2)},\hat{V}] & =-\sqrt{\frac{2}{3}}\frac{(N-2S)(N+2S+2)S(2S-1)}{(N-2)(N-3)}\\
Y_{1}^{(S,0)}[\hat{V}_{0},\hat{V}] & =2\frac{(N-1)(N-2S)(N+2S+2)S}{N-2}
\end{align}
\begin{align}
Y_{1}^{(S,0)}[\hat{V}_{-1}^{(2)},\hat{V}_{-1}^{(2)}] & =(N-1)(N-2S+2)(S-1)\frac{2(N+1)S^{2}-N(N+3)}{(N-2)(N-3)(S+1)}\\
Y_{2}^{(S,0)}[\hat{V}_{-1}^{(2)},\hat{V}_{-1}^{(2)}] & =\frac{(N-1)(N-2S)(N+2S)(N-2S+2)(S-1)}{2(N-2)(N-3)(S+1)}\\
Y_{1}^{(S,0)}[\hat{V}_{-1},\hat{V}_{-1}] & =N(N-1)(N-2S+2)\\
Y_{1}^{(S,0)}[\hat{V}_{-1}^{(2)},\hat{V}_{-1}] & =\frac{\sqrt{2}N(N-1)(N-2S+2)(S-1)}{(N-2)}\\
Y_{1}^{(S,0)}[\hat{V}_{-2}^{(2)},\hat{V}_{-2}^{(2)}] & =-\frac{(N-1)^{2}(N-2S+2)(N-2S+4)}{(N-2)(N-3)}\\
Y_{2}^{(S,0)}[\hat{V}_{-2}^{(2)},\hat{V}_{-2}^{(2)}] & =\frac{(N-1)(N-2S+2)(N-2S+4)}{2(N-3)}
\end{align}
The vanishing factors are omitted above, namely $Y_{2}^{(S,0)}[\hat{V}_{a},\hat{V}_{b}]=0$
when $\hat{V}_{a}$ or $\hat{V}_{b}$ is $\hat{V}_{k}$. The sum of
squared moduli of the matrix elements of the spin-independent interactions
(I.47b) can be expressed for bosons and the zero-range potentials
\refneq{VzeroRan} in the form \refneq{SumVaVbnn} too with
\begin{align}
Y_{1}^{(S,0)}[\hat{V},\hat{V}] & =-\frac{(N-1)(N-2S)(N+2S+2)}{(N-3)}\left(3-4S\frac{S+1}{N-2}\right)\\
Y_{2}^{(S,0)}[\hat{V},\hat{V}] & =(N-2S)(N+2S+2)\frac{3N(N-4)-4S(S+1)+12}{2(N-2)(N-3)}.
\end{align}
\end{subequations} \end{widetext}They are equal to the factors (I.47c).

Thus, sums of matrix elements and their squared moduli are expressed
in terms of universal factors, which are independent of the spatial
orbitals, and sums of one-body matrix elements (or their squared moduli),
which are independent of many-body spins and the spin-dependence of
the interaction. The sum rules, combined with the spin-projection
dependencies \refneq{AvectWE} and \refneq{VtensWE}, provide information
on each matrix element for any two-body spin-dependent interaction
between the particles.

\section{Multiplet energies and correlations for weakly-interacting gases\label{SecEnerCorr}}

\subsection{Average multiplet energies and energy widths}

Consider a general spin-dependent two-body interaction, conserving
the particle spins,
\begin{equation}
\hat{V}_{\mathrm{tot}}=\frac{1}{2}g_{\upuparrows}\hat{V}_{\upuparrows}+\frac{1}{2}g_{\downdownarrows}\hat{V}_{\downdownarrows}+g_{\uparrow\downarrow}\hat{V}_{\uparrow\downarrow},\label{Vtot}
\end{equation}
where the potentials are defined by Eqs. \refneq{Vupup}, \refneq{Vdowndown},
and \refneq{Vupdown} with the zero-range potential function \refneq{VzeroRan}.
Here the interaction strengths $g_{\sigma\sigma'}$ are proportional
to the $s$-wave elastic scattering lengthes $a_{\sigma\sigma'}$
for corresponding pairs of spin states. For example, in three-dimensional
geometry $g_{\sigma\sigma'}=4\pi\hbar^{2}a_{\sigma\sigma'}/m$, where
$m$ is the particle's mass. The factors $\frac{1}{2}$ in \refeq{Vtot}
appears due to double-counting of the interacting pairs in $\hat{V}_{\upuparrows}$
\refneq{Vupup} and $\hat{V}_{\downdownarrows}$ \refneq{Vdowndown}. 

In the case of weak interaction, the average multiplet energy can
be evaluated in the zero-order of the degenerate perturbation theory,
in the same way as in the case of spin-independent interactions {[}see
derivation of Eq. (I.52){]}
\begin{equation}
\bar{E}_{SS_{z}}=\frac{1}{f_{S}}\sum_{r}\langle\tilde{\Psi}_{r\{n\}S_{z}}^{(S)}|\hat{V}_{\mathrm{tot}}|\tilde{\Psi}_{r\{n\}S_{z}}^{(S)}\rangle\label{EbarSSz}
\end{equation}
 (in the case of spin-dependent interactions the energies depend on
the total spin projection $S_{z}$). 

For bosons with zero-range interactions, matrix elements of the potentials
are related to ones of the irreducible spherical tensor components
by Eqs. \refneq{Vupup}, \refneq{Vdowndown}, \refneq{Vupdown}, and
\refneq{SphVecTens0},
\[
\begin{aligned}\langle\Psi'|\hat{V}_{\upuparrows}|\Psi\rangle= & \langle\Psi'|\sqrt{\frac{2}{3}}\hat{V}_{0}^{(2)}+\hat{V}_{0}+\frac{1}{3}\hat{V}|\Psi\rangle\\
\langle\Psi'|\hat{V}_{\downdownarrows}|\Psi\rangle= & \langle\Psi'|\sqrt{\frac{2}{3}}\hat{V}_{0}^{(2)}-\hat{V}_{0}+\frac{1}{3}\hat{V}|\Psi\rangle\\
\langle\Psi'|\hat{V}_{\uparrow\downarrow}|\Psi\rangle= & \langle\Psi'|-\sqrt{\frac{2}{3}}\hat{V}_{0}^{(2)}+\frac{1}{6}\hat{V}|\Psi\rangle.
\end{aligned}
\]
These equations allow to expand the interaction $\hat{V}_{\mathrm{tot}}$
in terms of irreducible spherical tensors, which matrix elements can
be related to the ones for the maximal allowed spin projections by
Eqs. \refneq{AvectWE} and \refneq{VtensWE}. Then the sum rules \refneq{Y1S2}
lead to
\begin{multline*}
\bar{E}_{SS_{z}}=\frac{1}{2}\biggl(gY^{(S)}[\hat{V}]+g_{-}X_{S_{z}0}^{(S,S,1)}Y^{(S)}[\hat{V}_{0}]\\
+\sqrt{\frac{2}{3}}g_{+}X_{S_{z}0}^{(S,S,2)}Y^{(S)}[\hat{V}_{0}^{(2)}]\biggr)\langle V\rangle,
\end{multline*}
where $g=(g_{\upuparrows}+g_{\downdownarrows}+g_{\uparrow\downarrow})/3$,
$g_{+}=g_{\upuparrows}+g_{\downdownarrows}-2g_{\uparrow\downarrow}$,
$g_{-}=g_{\upuparrows}-g_{\downdownarrows}$, and the average matrix
element $\langle V\rangle$ is defined by \refeq{Vaver}. Substituting
the coefficients $X$ and $Y$ from Table 1 in \cite{yurovsky2015}
, Table \ref{TabVtens}, and \refeq{Y1S2} one gets
\begin{multline}
\bar{E}_{SS_{z}}=g\biggl[\frac{3}{4}N(N-2)+S(S+1)-\frac{1}{3}S(S+1)\frac{g_{+}}{g}\\
+S_{z}(N-1)\frac{g_{-}}{g}+S_{z}^{2}\frac{g_{+}}{g}\biggr]\langle V\rangle.\label{barESSz}
\end{multline}
Here the first two terms in the square brackets provide the average
multiplet energy (I.52) for spin-independent interactions. Spin-dependence
of the interactions leads the third term, which is independent of
the total spin projection $S_{z}$, as well as to the linear and quadratic
in $S_{z}$ shifts (the fourth and fifth terms, respectively). The
corrections are proportional to the ratios $g_{\pm}/g$, which are
determined by the scattering lengths. For example, $g_{+}/g\approx-0.001$
and $g_{-}/g\approx-0.049$ for $^{87}\mathrm{Rb}$. In this case,
the two states, generally used in experiments, $|\uparrow\rangle=|F=2,m_{f}=-1\rangle$
and $|\downarrow\rangle=|F=1,m_{f}=1\rangle$, have the scattering
lengths \cite{egorov2011} $a_{\upuparrows}\approx95.5a_{B}$, $a_{\downdownarrows}\approx100.4a_{B}$,
and $a_{\uparrow\downarrow}\approx98.0a_{B}$, where $a_{B}$ is the
Bohr radius. One-body spin-dependent interactions with external fields
lead only to linear shifts {[}see Eq. (I.54){]}.

The root-mean-square multiplet width can be evaluated in the same
way as in the case of spin-independent interactions {[}see derivation
of Eq. (I.53){]}
\[
\langle\Delta E_{SS_{z}}\rangle^{2}=\frac{1}{f_{S}}\sum_{r,r'}|\langle\tilde{\Psi}_{r'\{n\}S_{z}}^{(S)}|\hat{V}_{\mathrm{tot}}|\tilde{\Psi}_{r\{n\}S_{z}}^{(S)}\rangle|^{2}-\bar{E}_{SS_{z}}^{2}.
\]
Expanding the interaction $\hat{V}_{\mathrm{tot}}$ in terms of irreducible
spherical tensors, expressing their matrix elements in terms of the
ones for the maximal allowed spin projections, and applying the sum
rules \refneq{Y2SSp2nn} we get\begin{widetext}
\begin{multline*}
\langle\Delta E_{SS_{z}}\rangle^{2}=\frac{1}{4}\sum_{i=1}^{2}\biggl[g^{2}Y_{i}^{(S,0)}[\hat{V},\hat{V}]+\left(g_{-}X_{S_{z}0}^{(S,S,1)}\right)^{2}Y_{i}^{(S,0)}[\hat{V}_{0},\hat{V}_{0}]+\frac{2}{3}\left(g_{+}X_{S_{z}0}^{(S,S,2)}\right)^{2}Y_{i}^{(S,0)}[\hat{V}_{0}^{(2)},\hat{V}_{0}^{(2)}]\\
+2g_{-}gX_{S_{z}0}^{(S,S,1)}Y_{i}^{(S,0)}[\hat{V}_{0},\hat{V}]+2\sqrt{\frac{2}{3}}g_{+}gX_{S_{z}0}^{(S,S,2)}Y_{i}^{(S,0)}[\hat{V}_{0}^{(2)},\hat{V}]+2\sqrt{\frac{2}{3}}g_{+}g_{-}X_{S_{z}0}^{(S,S,1)}X_{S_{z}0}^{(S,S,2)}Y_{i}^{(S,0)}[\hat{V}_{0}^{(2)},\hat{V}_{0}]\biggr]\langle\Delta_{i}V\rangle^{2}.
\end{multline*}
Here the matrix element deviations $\langle\Delta_{i}V\rangle^{2}$
are defined by \refeq{DeltaV} and the terms proportional to $\langle V\rangle^{2}$
are canceled due to relation \refneq{Y20SS2nn}. The first term in
the square brackets gives the width for the spin-independent interactions
(I.53). Corrections due to spin-dependence of the interactions are
proportional to the small parameters $g_{\pm}/g$. Leading terms in
the coefficients before each power of $S_{z}$ can be obtained using
explicit expressions for $X$ from Table 1 in \cite{yurovsky2015}
 and Table \ref{TabVtens}, 
\begin{multline}
\langle\Delta E_{SS_{z}}\rangle^{2}=\frac{g^{2}}{4}\sum_{i=1}^{2}\biggl[Y_{i}^{(S,0)}[\hat{V},\hat{V}]-2\sqrt{\frac{2}{3}}\frac{S+1}{2S-1}Y_{i}^{(S,0)}[\hat{V}_{0}^{(2)},\hat{V}]\frac{g_{+}}{g}+\frac{2}{S}Y_{i}^{(S,0)}[\hat{V}_{0},\hat{V}]\frac{g_{-}}{g}S_{z}\\
+\frac{2\sqrt{6}}{S(2S-1)}Y_{i}^{(S,0)}[\hat{V}_{0}^{(2)},\hat{V}]\frac{g_{+}}{g}S_{z}^{2}+\frac{2\sqrt{6}}{S^{2}(2S-1)}Y_{i}^{(S,0)}[\hat{V}_{0}^{(2)},\hat{V}_{0}]\frac{g_{-}g_{+}}{g^{2}}S_{z}^{3}+\frac{6}{S^{2}(2S-1)^{2}}Y_{i}^{(S,0)}[\hat{V}_{0}^{(2)},\hat{V}_{0}^{(2)}]\frac{g_{+}^{2}}{g^{2}}S_{z}^{4}\biggr]\langle\Delta_{i}V\rangle^{2}.\label{DeltaESSz2}
\end{multline}

\end{widetext}

The consideration above was devoted to the case of bosons. For fermions,
matrix elements of $\hat{V}_{\upuparrows}$, and $\hat{V}_{\downdownarrows}$
vanish, according to the Pauli principle (see Sec. \ref{SecZeroRange}).
Then, due to Eqs. \refneq{Vupdown} and \refneq{V00Ferm} matrix elements
of $\hat{V}_{\mathrm{tot}}$ are equal to ones of $g_{\uparrow\downarrow}\hat{V}$.
Therefore, the muliplet average energies and energy widths are independent
of $S_{z}$ and can be calculated using Eqs. (I.52) and (I.53) for
spin-independent interactions.

\subsection{Average correlations}

The probabilities of finding two particles with given (either equal
or different) spins in the same point, the two-body local spin-dependent
correlations are expectation values of operators
\begin{equation}
\hat{\rho}_{\sigma_{1}\sigma_{2}}=\delta(\mathbf{r}_{1}-\mathbf{r}_{2})|\sigma_{1}(1)\rangle|\sigma_{2}(2)\rangle\langle\sigma_{1}(1)|\langle\sigma_{2}(2)|.\label{corr_oper}
\end{equation}
The spin projection $\sigma_{j}$ can be either $\uparrow$ or $\downarrow$.
Due to permutation symmetry of the total wavefunctions, the expectation
values of $\hat{\rho}_{\sigma_{1}\sigma_{2}}$ are proportional to
matrix elements of the spin-dependent potentials $\hat{V}_{\sigma_{1}\sigma_{2}}$
with the potential function \refneq{VzeroRan}. The multiplet-averaged
correlations can be evaluated in the same way as the average multiplet
energy \refneq{EbarSSz}. However, \refeq{EbarSSz} already contains
all necessary information, as, according to the Hellmann-Feinman theorem
\cite{hellmann1933,feynman1939}, the correlations are proportional
to derivatives of the average energy over respective coupling constants
\begin{multline*}
\bar{\rho}_{\sigma_{1}\sigma_{2}}^{(S,S_{z})}=\frac{1}{f_{S}}\sum_{r}\langle\tilde{\Psi}_{r\{n\}S_{z}}^{(S)}|\hat{\rho}_{\sigma_{1}\sigma_{2}}|\tilde{\Psi}_{r\{n\}S_{z}}^{(S)}\rangle\\
=\frac{1+\delta_{\sigma_{1}\sigma_{2}}}{N(N-1)}\frac{\partial}{\partial g_{\sigma_{1}\sigma_{2}}}\bar{E}_{SS_{z}}.
\end{multline*}
Then the dependence of the multiplet-averaged correlations
\begin{align*}
\bar{\rho}_{\upuparrows}^{(S,S_{z})}= & \biggl(\frac{1}{3}Y^{(S)}[\hat{V}]+X_{S_{z}0}^{(S,S,1)}Y^{(S)}[\hat{V}_{0}]\\
 & +\sqrt{\frac{2}{3}}X_{S_{z}0}^{(S,S,2)}Y^{(S)}[\hat{V}_{0}^{(2)}]\biggr)\frac{1}{N(N-1)}\langle\rho_{2}(0)\rangle\\
\bar{\rho}_{\downdownarrows}^{(S,S_{z})}= & \biggl(\frac{1}{3}Y^{(S)}[\hat{V}]-X_{S_{z}0}^{(S,S,1)}Y^{(S)}[\hat{V}_{0}]\\
 & +\sqrt{\frac{2}{3}}X_{S_{z}0}^{(S,S,2)}Y^{(S)}[\hat{V}_{0}^{(2)}]\biggr)\frac{1}{N(N-1)}\langle\rho_{2}(0)\rangle\\
\bar{\rho}_{\uparrow\downarrow}^{(S,S_{z})}= & \left(\frac{1}{6}Y^{(S)}[\hat{V}]-\sqrt{\frac{2}{3}}X_{S_{z}0}^{(S,S,2)}Y^{(S)}[\hat{V}_{0}^{(2)}]\right)\\
 & \times\frac{1}{N(N-1)}\langle\rho_{2}(0)\rangle
\end{align*}
on the total many-body spin $S$ and its projection $S_{z}$ is factorized
to the universal factors, which are independent of the spatial Hamiltonian
and occupied spatial orbitals. These factors are expressed in terms
of the coefficients $X$ and $Y$. The dependence on the spatial state
is given by the average two-body density \cite{yurovsky2014}, $\langle\rho_{2}(0)\rangle$,
which is independent of $S$ and $S_{z}$ and equal to the average
matrix element $\langle V\rangle$ \refneq{Vaver} for the potential
function \refneq{VzeroRan}, $\langle\rho_{2}(0)\rangle=\langle V\rangle$.
Similar factorization was proved \cite{yurovsky2014} for spin-independent
local correlations of particles with arbitrary spins. Substitution
the coefficients $X$ and $Y$ from Table 1 in \cite{yurovsky2015}
, Table \ref{TabVtens}, and \refeq{Y1S2} leads to explicit expressions
for the universal factors in terms of $S$ and $S_{z}$, 
\begin{align}
\bar{\rho}_{\upuparrows}^{(S,S_{z})}= & \frac{\langle\rho_{2}(0)\rangle}{N(N-1)}\left(\frac{1}{2}N(N-2)+2(N-1)S_{z}+2S_{z}^{2}\right)\nonumber \\
\bar{\rho}_{\downdownarrows}^{(S,S_{z})}= & \frac{\langle\rho_{2}(0)\rangle}{N(N-1)}\left(\frac{1}{2}N(N-2)-2(N-1)S_{z}+2S_{z}^{2}\right)\label{barrhoSSz}\\
\bar{\rho}_{\uparrow\downarrow}^{(S,S_{z})}= & \frac{\langle\rho_{2}(0)\rangle}{N(N-1)}\left(\frac{1}{4}N(N-2)+S(S+1)-2S_{z}^{2}\right)\nonumber 
\end{align}

The local spin-independent correlations \cite{yurovsky2014} are multiplet-averaged
expectation values of $\delta(\mathbf{r}_{1}-\mathbf{r}_{2})$ and
can be calculated with the characters from Table 2 in \cite{yurovsky2015}
 for both bosons and fermions (the signs $+$ and $-$, respectively,
below) 
\begin{align}
\bar{\rho}_{2}^{[\lambda]}(0) & =\left(1\pm\frac{\chi_{S}(\{2\})}{f_{S}}\right)\langle\rho_{2}(0)\rangle\nonumber \\
 & =\left(1\pm\frac{4S^{2}+N^{2}+4S-4N}{2N(N-1)}\right)\langle\rho_{2}(0)\rangle.\label{barrho2}
\end{align}
They are related to spin-dependent correlations, $\bar{\rho}_{2}^{[\lambda]}(0)=\bar{\rho}_{\upuparrows}^{(S,S_{z})}+\bar{\rho}_{\downdownarrows}^{(S,S_{z})}+2\bar{\rho}_{\uparrow\downarrow}^{(S,S_{z})}$,
as can be proved in the same way as \refneq{Vuduudd}, and are independent
of $S_{z}$.

In an alternative description, each particle is characterized by its
spin projection and coordinate, and the total wavefunction is symmetrized
for bosons or antisymmetrized for fermions over permutations of all
particles {[}see Eq. (I.19){]},
\begin{equation}
\tilde{\Psi}_{\{n\}\{\sigma\}}=(N!)^{-1/2}\sum_{\pr P}\mathrm{sgn}(\pr P)\prod_{j=1}^{N}\varphi_{n_{\pr Pj}}(\mathbf{r}_{j})|\sigma_{\pr Pj}(j)\rangle,\label{tilPsinsig}
\end{equation}
where the factor $\mathrm{sgn}(\pr P)$ is the permutation parity
for fermions and $\mathrm{sgn}(\pr P)\equiv1$ for bosons. Given total
spin projection $S_{z}$, the set $\{\sigma\}$ contains $N_{\uparrow}=N/2+S_{z}$
spins $\uparrow$ and $N_{\downarrow}=N/2-S_{z}$ spins $\downarrow$.
For these wavefunctions the correlations are calculated as expectation
values of the operators \refneq{corr_oper} and averaged over all
distinct choices of $N_{\uparrow}$ particles with spin up, leading
to

\begin{align*}
\bar{\rho}_{\upuparrows}^{(N_{\uparrow},N_{\downarrow})}= & \frac{1+\mathrm{sgn}(\pr P_{12})}{N(N-1)}N_{\uparrow}(N_{\uparrow}-1)\langle\rho_{2}(0)\rangle\\
\bar{\rho}_{\downdownarrows}^{(N_{\uparrow},N_{\downarrow})}= & \frac{1+\mathrm{sgn}(\pr P_{12})}{N(N-1)}N_{\downarrow}(N_{\downarrow}-1)\langle\rho_{2}(0)\rangle.\\
\bar{\rho}_{\uparrow\downarrow}^{(N_{\uparrow},N_{\downarrow})}= & \frac{\langle\rho_{2}(0)\rangle}{N(N-1)}N_{\uparrow}N_{\downarrow}
\end{align*}
The transposition parity $\mathrm{sgn}(\pr P_{12})=1$ leads to the
factor $2$ in $\bar{\rho}_{\upuparrows}^{(N_{\uparrow},N_{\downarrow})}$
and $\bar{\rho}_{\upuparrows}^{(N_{\uparrow},N_{\downarrow})}$ for
bosons. For fermions, $\mathrm{sgn}(\pr P_{12})=-1$ and correlations
of particles with the same spins vanish, according to the Pauli exclusion
principle. Using relations between $S_{z}$ and $N_{\uparrow\downarrow}$,
one can see that the average correlations of bosons with the same
spins are the same as for wavefunctions with defined collective spins
and individual spin projections, $\bar{\rho}_{\upuparrows}^{(S,S_{z})}=\bar{\rho}_{\upuparrows}^{(N_{\uparrow},N_{\downarrow})}$,
$\bar{\rho}_{\downdownarrows}^{(S,S_{z})}=\bar{\rho}_{\upuparrows}^{(N_{\uparrow},N_{\downarrow})}$.
However, average correlations of particles with opposite spins are
different, 
\[
\bar{\rho}_{\uparrow\downarrow}^{(S,S_{z})}=\bar{\rho}_{\uparrow\downarrow}^{(N_{\uparrow},N_{\downarrow})}+\frac{\langle\rho_{2}(0)\rangle}{N(N-1)}\left(S^{2}-S_{z}^{2}+S-\frac{N}{2}\right).
\]
The same is valid for average local spin-independent correlations,
calculated as a sum of spin-dependent correlations. For the defined
individual spin projections we have for both bosons and fermions (the
signs $+$ and $-$, respectively, below) 
\begin{multline*}
\bar{\rho}_{\upuparrows}^{(N_{\uparrow},N_{\downarrow})}+\bar{\rho}_{\downdownarrows}^{(N_{\uparrow},N_{\downarrow})}+2\bar{\rho}_{\uparrow\downarrow}^{(N_{\uparrow},N_{\downarrow})}\\
=\left(1\pm\frac{N^{2}+4S_{z}^{2}-2N}{2N(N-1)}\right)\langle\rho_{2}(0)\rangle,
\end{multline*}
which depends on the total spin projection, unlike $\bar{\rho}_{2}^{[\lambda]}(0)$
{[}see \refeq{barrho2}{]}. For fermions, the average correlations
of particles with opposite spins, being equal to a half spin-independent
ones, are different for the two kinds of states too. Thus the spin-independent
correlations, as well as the correlations of particles with opposite
spins, allow to determine the kind of the many-body state.

\section*{Conclusions}

Matrix elements of spin-dependent two-body interactions {[}Eqs. \refneq{TwoBodySpinConserv},
\refneq{TwoBodyOneSpinChange}, and \refneq{TwoBodyTwoSpinChange}{]}
in the basis with collective spin and spatial wavefunctions \refneq{tilPsiSrnSz}
can be calculated with group-theoretical methods. These matrix elements
agree to the selection rules \cite{yurovsky2014}. The interactions
can be decomposed into irreducible spherical tensors, whose explicit
dependencies on the total spin projection {[}Eqs. \refneq{AvectWE}
and \refneq{VtensWE}{]} are obtained using the Wigner-Eckart theorem.
Analytic expressions are derived for sums of these matrix elements
\refneq{Y1S2} and their squared moduli {[}Eqs. \refneq{Y2S2} and
\refneq{Y2SSp2nn}{]} over wavefunctions with the fixed total spin,
its projection, and the set of spatial orbitals. Dependence on the
many-body states in these sums is given by the $3j$ Wigner symbols
and the universal factors $Y^{(S)}$, $Y^{(S,2)}$, $Y_{0}^{(S,0)}$,
$Y_{1}^{(S,0)}$, and $Y_{2}^{(S,0)}$. These factors are independent
of details of one-body Hamiltonians and are expressed in terms of
the total spin and number of particles. The sum rules can be applied
to the evaluation of changes of the spin-multiplet average energies
\refneq{barESSz} and energy widths \refneq{DeltaESSz2} due to weak
spin-dependent interactions. Mutiplet-averaged two-body spin-dependent
correlations \refneq{barrhoSSz}, calculated with the sum rules, are
factorized to universal factors, which are independent of the spatial
orbitals, and the average density, which is independent of many-body
spins. The difference between these correlations and ones for the
many-body states with defined individual spin projections allows identification
of the many-body state kind. Other possible applications of the sum
rules include estimates of the spin-multiplet depletion rates due
to spin-dependent two-body perturbations.

\appendix

\section{Calculation of the sums \refneq{SigmaSSp4}\label{AppSigmaSSp}}

The sums \refneq{SigmaSSp4} and \refneq{SigmaS4} contain the Young
orthogonal matrix elements $D_{[0][0]}^{[\lambda]}(\pr R)$, which
have been calculated by Goddard \cite{goddard1967} in the following
way. Each permutation $\pr R$ can be represented as
\begin{equation}
\pr R=\prod_{k=1}^{n_{ex}}\pr P_{i'_{k}i''_{k}}\pr P'\pr P'',\label{PermTranPpPpp}
\end{equation}
 where $\pr P'$ are permutations of symbols in the first row of the
Young tableau $[0]$ ($\lambda_{1}$ first symbols), $\pr P''$ are
permutations of symbols in the second row ($\lambda_{2}$ last symbols),
and $\pr P_{i'_{k}i''_{k}}$ transpose symbols between the rows as
$i'_{k}\leq\lambda_{1}$ and $i''_{k}>\lambda_{1}$. Then \cite{goddard1967}
the matrix element is inversely proportional to the binomial coefficient,
\begin{equation}
D_{[0][0]}^{[\lambda]}(\pr R)=(-1)^{n_{ex}}\binom{\lambda_{1}}{n_{ex}}^{-1}=(-1)^{n_{ex}}\frac{n_{ex}!(\lambda_{1}-n_{ex})!}{\lambda_{1}!}.\label{D00Godd}
\end{equation}

Using relations \refneq{RepProd} and \refneq{InvOrthMat} and substitution
$\pr R=\pr Q\pr P^{-1}$, the sum \refneq{SigmaSSp4} can be represented
in the following form \begin{widetext}
\[
\varSigma_{j_{1}j'_{1}j_{2}j'_{2}}^{(S',S)}(l_{1},l'_{1},l_{2},l'_{2})=\sum_{\pr R}D_{[0][0]}^{[\lambda']}(\pr R)D_{[0][0]}^{[\lambda]}(\pr R)\sum_{\pr P}\delta_{l_{1},\pr Pj_{1}}\delta_{l'_{1},\pr Pj_{1}'}\delta_{l_{2},\pr{R\pr P}j_{2}}\delta_{l'_{2},\pr{R\pr P}j_{2}'},
\]
where $\lambda=[N/2+S,N/2-S]$ and $\lambda'=[N/2+S',N/2-S']$. Only
the sums with $j_{1}\neq j'_{1}$ and $j_{2}\neq j'_{2}$ are used
here. This implies $l_{1}\neq l'_{1}$ and $l_{2}\neq l'_{2}$. The
sum remains unchanged on simultaneous permutation of arguments and
corresponding subscripts, $\varSigma_{j_{1}j'_{1}j_{2}j'_{2}}^{(S',S)}(l_{1},l'_{1},l_{2},l'_{2})=\varSigma_{j'_{1}j_{1}j_{2}j'_{2}}^{(S',S)}(l'_{1},l_{1},l_{2},l'_{2})=\varSigma_{j_{1}j'_{1}j'_{2}j_{2}}^{(S',S)}(l_{1},l'_{1},l'_{2},l_{2})$.

If $j_{1}=j_{2}$ and $j'_{1}=j'_{2}$, 
\begin{equation}
\varSigma_{jj'jj'}^{(S',S)}(l_{1},l'_{1},l_{2},l'_{2})=(N-2)!\sum_{\pr R}D_{[0][0]}^{[\lambda']}(\pr R)D_{[0][0]}^{[\lambda]}(\pr R)\delta_{l_{2},\pr Rl_{1}}\delta_{l'_{2},\pr Rl'_{1}},\label{SigmaSSpjjpjjp}
\end{equation}
since there are $(N-2)!$ permutations $\pr P$ such that $l_{1}=\pr Pj_{1}$
and $l'_{1}=\pr Pj_{1}'$. This sum is independent of $j$ and $j'$.
Due to the invariance mentioned above, $\varSigma_{jj'j'j}^{(S',S)}(l_{1},l'_{1},l_{2},l'_{2})=\varSigma_{jj'jj'}^{(S',S)}(l_{1},l'_{1},l'_{2},l_{2})$
and $\varSigma_{jj'jj'}^{(S',S)}(l'_{1},l_{1},l'_{2},l_{2})=\varSigma_{jj'jj'}^{(S',S)}(l_{1},l'_{1},l_{2},l'_{2})$.
The identity $\varSigma_{jj'jj'}^{(S,S')}(l_{1},l'_{1},l_{2},l'_{2})=\varSigma_{jj'jj'}^{(S,S')}(l_{2},l'_{2},l_{1},l'_{1})$
can be proved by the substitution $\pr R=\pr R^{-1}$.

If $j'_{1}=j'_{2}$, but $j_{1}\neq j_{2}$, we have
\[
\sum_{\pr P}\delta_{l_{1},\pr Pj_{1}}\delta_{l'_{1},\pr Pj'}\delta_{l_{2},\pr{R\pr P}j_{2}}\delta_{l'_{2},\pr{R\pr P}j'}=\delta_{l'_{2},\pr Rl'_{1}}\sum_{l_{1}\neq l\neq l'_{1}}\delta_{l_{2},\pr Rl}\sum_{\pr P}\delta_{l_{1},\pr Pj_{1}}\delta_{l,\pr Pj_{2}}\delta_{l'_{1},\pr Pj'}=(N-3)!\delta_{l'_{2},\pr Rl'_{1}}\sum_{l}\delta_{l_{2},\pr Rl}(1-\delta_{ll_{1}}-\delta_{ll'_{1}}).
\]
Then
\begin{multline*}
\varSigma_{j_{1}j'j_{2}j'}^{(S',S)}(l_{1},l'_{1},l_{2},l'_{2})=\sum_{\pr R}D_{[0][0]}^{[\lambda']}(\pr R)D_{[0][0]}^{[\lambda]}(\pr R)(N-3)!\delta_{l'_{2},\pr Rl'_{1}}(1-\delta_{l_{2},\pr Rl_{1}}-\delta_{l_{2},\pr Rl'_{1}})\\
=\frac{1}{(N-1)(N-2)}\varSigma_{jj}^{(S',S)}(l'_{1},l'_{2})-\frac{1}{N-2}\varSigma_{jj'jj'}^{(S',S)}(l_{1},l'_{1},l_{2},l'_{2})
\end{multline*}
($\delta_{l'_{2},\pr Rl'_{1}}\delta_{l_{2},\pr Rl'_{1}}=0,$ since
$l_{2}\neq l'_{2}$) is independent of $j'$ and $j_{1}\neq j_{2}$.
Here the sum 
\begin{equation}
\varSigma_{jj}^{(S',S)}(l,l')=(N-1)!\sum_{\pr R}D_{[0][0]}^{[\lambda']}(\pr R)D_{[0][0]}^{[\lambda]}(\pr R)\delta_{l',\pr Rl}\label{SigmaSSpjj}
\end{equation}
is calculated in Appendix \ref{AppSSpjj}. The sum 
\[
\varSigma_{3}^{(S',S)}(l_{1},l'_{1},l_{2},l'_{2})=\varSigma_{j'j_{1}j'j_{2}}^{(S',S)}(l_{1},l'_{1},l_{2},l'_{2})+\varSigma_{j'j_{1}j_{2}j'}^{(S',S)}(l_{1},l'_{1},l_{2},l'_{2})+\varSigma_{j_{1}j'j'j_{2}}^{(S',S)}(l_{1},l'_{1},l_{2},l'_{2})+\varSigma_{j_{1}j'j_{2}j'}^{(S',S)}(l_{1},l'_{1},l_{2},l'_{2})
\]
 in \refeq{SumSigVV} can be expressed as
\begin{equation}
\varSigma_{3}^{(S',S)}(l_{1},l'_{1},l_{2},l'_{2})=\frac{1}{(N-1)(N-2)}\varSigma_{1}^{(S',S)}(l_{1},l'_{1},l_{2},l'_{2})-\frac{2}{N-2}\varSigma_{2}^{(S',S)}(l_{1},l'_{1},l_{2},l'_{2}),
\end{equation}
where
\begin{equation}
\varSigma_{1}^{(S',S)}(l_{1},l'_{1},l_{2},l'_{2})=\varSigma_{jj}^{(S',S)}(l_{1},l_{2})+\varSigma_{jj}^{(S',S)}(l'_{1},l_{2})+\varSigma_{jj}^{(S',S)}(l_{1},l'_{2})+\varSigma_{jj}^{(S',S)}(l'_{1},l'_{2})\label{Sigma1}
\end{equation}
and $\varSigma_{2}^{(S',S)}(l_{1},l'_{1},l_{2},l'_{2})$ is defined
by \refeq{SigmaSSp2}.

Finally, if $j_{1}\neq j_{2}\neq j'_{1}$ and $j_{1}\neq j'_{2}\neq j_{2}$,
the sum is expressed using the following identity, 
\begin{multline*}
\sum_{\pr P}\delta_{l_{1},\pr Pj_{1}}\delta_{l'_{1},\pr Pj'_{1}}\delta_{l_{2},\pr{R\pr P}j_{2}}\delta_{l'_{2},\pr{R\pr P}j'_{2}}=\sum_{l_{1}\neq l\neq l'_{1}}\delta_{l_{2},\pr Rl}\sum_{l_{1}\neq l'\neq l'_{1}}\delta_{l'_{2},\pr Rl'}\sum_{\pr P}\delta_{l_{1},\pr Pj_{1}}\delta_{l'_{1},\pr Pj'_{1}}\delta_{l,\pr Pj_{2}}\delta_{l',\pr Pj'_{2}}\\
=(N-4)!\sum_{l,l'}\delta_{l_{2},\pr Rl}\delta_{l'_{2},\pr Rl'}(1-\delta_{ll_{1}})(1-\delta_{ll'_{1}})(1-\delta_{l'l_{1}})(1-\delta_{l'l'_{1}}).
\end{multline*}
Then $\varSigma_{j_{1}j'_{1}j_{2}j'_{2}}^{(S',S)}(l_{1},l'_{1},l_{2},l'_{2})\equiv\varSigma_{4}^{(S',S)}(l_{1},l'_{1},l_{2},l'_{2})$
{[}see \refeq{SumSigVV}{]} is independent of $j_{1}$, $j_{2}$,
$j'_{1}$, and $j'_{2}$ and 
\begin{multline}
\varSigma_{4}^{(S',S)}(l_{1},l'_{1},l_{2},l'_{2})=\sum_{\pr R}D_{[0][0]}^{[\lambda']}(\pr R)D_{[0][0]}^{[\lambda]}(\pr R)(N-4)!(1-\delta_{l_{2},\pr Rl_{1}}-\delta_{l_{2},\pr Rl'_{1}}-\delta_{l'_{2},\pr Rl_{1}}-\delta_{l'_{2},\pr Rl'_{1}}+\delta_{l_{2},\pr Rl_{1}}\delta_{l'_{2},\pr Rl'_{1}}+\delta_{l_{2},\pr Rl'_{1}}\delta_{l'_{2},\pr Rl_{1}})\\
=\frac{N!(N-4)!}{f_{S}}\delta_{\lambda\lambda'}-\frac{1}{(N-1)(N-2)(N-3)}\varSigma_{1}^{(S',S)}(l_{1},l'_{1},l_{2},l'_{2})+\frac{1}{(N-2)(N-3)}\varSigma_{2}^{(S',S)}(l_{1},l'_{1},l_{2},l'_{2}).
\end{multline}
{[}see Eqs. (I.5), \refneq{SigmaSSp2} and \refneq{Sigma1}{]}. 

Therefore, each sum \refneq{SigmaSSp4} is expressed in terms of \refneq{SigmaSSpjjpjjp}
and \refneq{SigmaSSpjj}.

Let us at first calculate the sums \refneq{SigmaSSpjjpjjp} for $S=S'$.
If $l_{1}=l_{2}=\lambda_{1}-1$, $l'_{1}=l'_{2}=\lambda_{1}$, the
Kronecker symbols in \refeq{SigmaSSpjjpjjp} select the permutations
of the form \refneq{PermTranPpPpp} with $\pr P'$ which do not affect
$l_{1}$ and $l'_{1}$ and $l_{1}\neq i'_{k}\neq l'_{1}$. Therefore
there are $(\lambda_{1}-2)!$ permutations $\pr P'$ , $\lambda_{2}!$
permutations $\pr P''$, and number of distinct choices of the sets
of $i'_{k}$ and $i''_{k}$ are given by the binomial coefficients
$\binom{\lambda_{1}-2}{n_{ex}}$ and $\binom{\lambda_{2}}{n_{ex}}$
, respectively. Then \refeq{SigmaSSpjjpjjp} can be transformed as
follows,
\begin{multline*}
\varSigma_{jj'jj'}^{(S,S)}(\lambda_{1}-1,\lambda_{1},\lambda_{1}-1,\lambda_{1})=(N-2)!\sum_{n_{ex}=0}^{\lambda_{2}}(\lambda_{1}-2)!\lambda_{2}!\binom{\lambda_{1}-2}{n_{ex}}\binom{\lambda_{2}}{n_{ex}}\binom{\lambda_{1}}{n_{ex}}^{-2}\\
=\frac{N!(N-2)!}{f_{S}\lambda_{1}(\lambda_{1}-1)}\left[1-\frac{2\lambda_{2}}{(\lambda_{1}-1)(\lambda_{1}-\lambda_{2}+3)}+\frac{4\lambda_{2}}{\lambda_{1}(\lambda_{1}-1)(\lambda_{1}-\lambda_{2}+2)(\lambda_{1}-\lambda_{2}+3)}\right].
\end{multline*}

If $l_{1}=l'_{2}=\lambda_{1}-1$, $l'_{1}=l_{2}=\lambda_{1}$, the
permutations
\begin{equation}
\pr R=\pr P_{l_{1}l_{2}}\prod_{k=1}^{k_{m}}\pr P_{i'_{k}i''_{k}}\pr P'\pr P''\label{Pl1l2TranPpPpp}
\end{equation}
satisfy the Kronecker symbols if $\pr P'$ do not affect $l_{1}$
and $l'_{1}$ and $l_{1}\neq i'_{k}\neq l'_{1}$. Since $\pr P_{l_{1}l_{2}}$
is a transposition of symbols in the first row of the Young tableau
$[0]$ , $n_{ex}=k_{m}$ and 

\[
\varSigma_{jj'jj'}^{(S,S)}(\lambda_{1}-1,\lambda_{1},\lambda_{1},\lambda_{1}-1)=\varSigma_{jj'jj'}^{(S,S)}(\lambda_{1}-1,\lambda_{1},\lambda_{1}-1,\lambda_{1})
\]

If $l_{1}=l_{2}=\lambda_{1}+1$, $l'_{1}=l'_{2}=\lambda_{1}+2$, the
proper permutations are given by \refneq{PermTranPpPpp} with $\pr P''$
which do not affect $l_{1}$ and $l'_{1}$ and $l_{1}\neq i''_{k}\neq l'_{1}$,
and with no additional restrictions to $\pr P'$ and $i'_{k}$ . There
are $\lambda_{1}!$ permutations $\pr P'$ , $(\lambda_{2}-2)!$ permutations
$\pr P''$, and number of distinct choices of the sets of $i'_{k}$
and $i''_{k}$ are given by the binomial coefficients $\binom{\lambda_{1}}{n_{ex}}$
and $\binom{\lambda_{2}-2}{n_{ex}}$ , respectively. Then
\begin{multline*}
\varSigma_{jj'jj'}^{(S,S)}(\lambda_{1}+1,\lambda_{1}+2,\lambda_{1}+1,\lambda_{1}+2)=(N-2)!\sum_{n_{ex}=0}^{\lambda_{2}}\lambda_{1}!(\lambda_{2}-2)!\binom{\lambda_{1}}{n_{ex}}\binom{\lambda_{2}-2}{n_{ex}}\binom{\lambda_{1}}{n_{ex}}^{-2}\\
=\frac{N!(N-2)!}{f_{S}\lambda_{2}(\lambda_{2}-1)}\left[1-\frac{2}{\lambda_{1}-\lambda_{2}+3}\right].
\end{multline*}
If $l_{1}=l'_{2}=\lambda_{1}+1$, $l'_{1}=l_{2}=\lambda_{1}+2$, the
Kronecker symbols are satisfied by the permutations \refneq{Pl1l2TranPpPpp}
with the same restrictions to $\pr P''$ and $i''_{k}$ as in the
previous case. Since $\pr P_{l_{1}l_{2}}$ is a transposition of symbols
in the second row of the Young tableau $[0]$ , $n_{ex}=k_{m}$ and 

\[
\varSigma_{jj'jj'}^{(S,S)}(\lambda_{1}+1,\lambda_{1}+2,\lambda_{1}+2,\lambda_{1}+1)=\varSigma_{jj'jj'}^{(S,S)}(\lambda_{1}+1,\lambda_{1}+2,\lambda_{1}+1,\lambda_{1}+2).
\]

If $l_{1}=\lambda_{1}-1$, $l'_{1}=\lambda_{1}$, $l_{2}=\lambda_{1}+1$,
and $l'_{2}=\lambda_{1}+2$, the permutations 
\begin{equation}
\pr R=\pr P_{l_{1}l_{2}}\pr P_{l'_{1}l'_{2}}\prod_{k=1}^{k_{m}}\pr P_{i'_{k}i''_{k}}\pr P'\pr P''\label{Pl1l2Pl1pl2pTranPpPpp}
\end{equation}
satisfy the Kronecker symbol if $\pr P'$ do not affect $l_{1}$ and
$l'_{1}$ and $l_{1}\neq i'_{k}\neq l'_{1}$. If $l_{2}\neq i''_{k}\neq l'_{2}$
for any $k$, $\pr P_{l_{1}l_{2}}$and $\pr P_{l'_{1}l'_{2}}$are
additional traspositions between the rows of the Young tableau $[0]$,
and $n_{ex}=k_{m}+2$. Otherwise, if one of $i''_{k}$ is equal to
$l_{2}$ , $\pr P_{l_{1}l_{2}}\pr P_{i'_{k}l_{2}}=\pr P_{i'_{k}l_{2}}\pr P_{i'_{k}l_{1}}$,
$n_{ex}=k_{m}+1$. By the same reason $n_{ex}=k_{m}+1$ if one of
$i''_{k}$ is equal to $l'_{2}$ . If two of $i''_{k}$ are equal
to $l_{2}$ and $l'_{2}$, $n_{ex}=k_{m}$. Then
\begin{multline*}
\varSigma_{jj'jj'}^{(S,S)}(\lambda_{1}-1,\lambda_{1},\lambda_{1}+1,\lambda_{1}+2)=(N-2)!\sum_{k_{m}=0}^{\lambda_{2}}(\lambda_{1}-2)!\lambda_{2}!\binom{\lambda_{1}-2}{k_{m}}\biggl[\binom{\lambda_{2}-2}{k_{m}}\binom{\lambda_{1}}{k_{m}+2}^{-2}+2\binom{\lambda_{2}-2}{k_{m}-1}\binom{\lambda_{1}}{k_{m}+1}^{-2}\\
+\binom{\lambda_{2}-2}{k_{m}-2}\binom{\lambda_{1}}{k_{m}}^{-2}\biggr]=\frac{2N!(N-2)!}{f_{S}\lambda_{1}(\lambda_{1}-1)(\lambda_{1}-\lambda_{2}+2)(\lambda_{1}-\lambda_{2}+3)}.
\end{multline*}
This derivation is valid for the case of $l_{1}=\lambda_{1}-1$, $l'_{1}=\lambda_{1}$,
$l_{2}=\lambda_{1}+2$, and $l'_{2}=\lambda_{1}+1$ as well, giving
\[
\varSigma_{jj'jj'}^{(S,S)}(\lambda_{1}-1,\lambda_{1},\lambda_{1}+2,\lambda_{1}+1)=\varSigma_{jj'jj'}^{(S,S)}(\lambda_{1}-1,\lambda_{1},\lambda_{1}+1,\lambda_{1}+2).
\]

Consider now the case of $S'=S-1$. If $l_{1}=l_{2}=\lambda_{1}-1$,
$l'_{1}=l'_{2}=\lambda_{1}$, the Kronecker symbols in \refeq{SigmaSSpjjpjjp}
select the permutations of the form \refneq{PermTranPpPpp} with $\pr P'$
which do not affect $l_{1}$ and $l'_{1}$ and $l_{1}\neq i'_{k}\neq l'_{1}$.
Then
\begin{multline*}
\varSigma_{jj'jj'}^{(S-1,S)}(\lambda_{1}-1,\lambda_{1},\lambda_{1}-1,\lambda_{1})=(N-2)!\sum_{n_{ex}=0}^{\lambda_{2}}(\lambda_{1}-2)!\lambda_{2}!\binom{\lambda_{1}-2}{n_{ex}}\binom{\lambda_{2}}{n_{ex}}\binom{\lambda_{1}}{n_{ex}}^{-1}\binom{\lambda_{1}-1}{n_{ex}}^{-1}\\
=\frac{N!(N-2)!(\lambda_{1}^{2}-\lambda_{1}\lambda_{2}+\lambda_{1}-2)}{f_{S}\lambda_{1}(\lambda_{1}-1)^{2}(\lambda_{1}-\lambda_{2}+2)}.
\end{multline*}

If $l_{1}=l'_{2}=\lambda_{1}-1$, $l'_{1}=l_{2}=\lambda_{1}$, the
permutations \refneq{Pl1l2TranPpPpp} satisfy the Kronecker symbols
if $\pr P'$ do not affect $l_{1}$ and $l'_{1}$ and $l_{1}\neq i'_{k}\neq l'_{1}$.
Since $\pr P_{l_{1}l_{2}}$ is an additional transposition between
the rows of the Young tableau $[0]$ of the shape $\lambda'=[\lambda_{1}-1,\lambda_{2}+1]$,
$n_{ex}=k_{m}$ and $n'_{ex}=k_{m}+1$. As a result,
\begin{multline*}
\varSigma_{jj'jj'}^{(S-1,S)}(\lambda_{1}-1,\lambda_{1},\lambda_{1},\lambda_{1}-1)=-(N-2)!\sum_{k_{m}=0}^{\lambda_{2}}(\lambda_{1}-2)!\lambda_{2}!\binom{\lambda_{1}-2}{k_{m}}\binom{\lambda_{2}}{k_{m}}\binom{\lambda_{1}}{k_{m}}^{-1}\binom{\lambda_{1}-1}{k_{m}+1}^{-1}\\
=-\frac{N!(N-2)!(\lambda_{1}+2)}{f_{S}\lambda_{1}(\lambda_{1}-1)^{2}(\lambda_{1}-\lambda_{2}+2)}.
\end{multline*}

If $l_{1}=l_{2}=\lambda_{1}$, $l'_{1}=l'_{2}=\lambda_{1}+1$, the
Kronecker symbols in \refeq{SigmaSSpjjpjjp} are satisfied by the
permutations of the form \refneq{PermTranPpPpp} with $\pr P'$ which
do not affect $l_{1}$ , $\pr P''$ which do not affect $l'_{1}$,
$i'_{k}\neq l_{1}$, and $i''_{k}\neq l'_{1}$. Then
\begin{multline*}
\varSigma_{jj'jj'}^{(S-1,S)}(\lambda_{1},\lambda_{1}+1,\lambda_{1},\lambda_{1}+1)=(N-2)!\sum_{n_{ex}=0}^{\lambda_{2}}(\lambda_{1}-1)!(\lambda_{2}-1)!\binom{\lambda_{1}-1}{n_{ex}}\binom{\lambda_{2}-1}{n_{ex}}\binom{\lambda_{1}}{n_{ex}}^{-1}\binom{\lambda_{1}-1}{n_{ex}}^{-1}\\
=\frac{N!(N-2)!(\lambda_{1}-\lambda_{2}+1)}{f_{S}\lambda_{1}\lambda_{2}(\lambda_{1}-\lambda_{2}+2)}.
\end{multline*}

If $l_{1}=l'_{2}=\lambda_{1}$, $l'_{1}=l_{2}=\lambda_{1}+1$, the
permutations \refneq{Pl1l2TranPpPpp} satisfy the Kronecker symbols
if $\pr P'$ do not affect $l_{1}$, $\pr P''$ do not affect $l'_{1}$,
$i'_{k}\neq l_{1}$, and $i''_{k}\neq l'_{1}$. Since $\pr P_{l_{1}l_{2}}$
is an additional transposition between the rows of the Young tableau
$[0]$ of the shape $\lambda$, $n_{ex}=k_{m}+1$ and $n'_{ex}=k_{m}$.
As a result,
\begin{multline*}
\varSigma_{jj'jj'}^{(S-1,S)}(\lambda_{1},\lambda_{1}+1,\lambda_{1}+1,\lambda_{1})=-(N-2)!\sum_{k_{m}=0}^{\lambda_{2}}(\lambda_{1}-1)!(\lambda_{2}-1)!\binom{\lambda_{1}-1}{k_{m}}\binom{\lambda_{2}-1}{k_{m}}\binom{\lambda_{1}}{k_{m}+1}^{-1}\binom{\lambda_{1}-1}{k_{m}}^{-1}\\
=-\frac{N!(N-2)!}{f_{S}\lambda_{1}\lambda_{2}(\lambda_{1}-\lambda_{2}+2)}.
\end{multline*}

If $l_{1}=\lambda_{1}-1$, $l'_{1}=l'_{2}=\lambda_{1}$, $l_{2}=\lambda_{1}+1$,
the proper permutations are given by \refneq{Pl1l2TranPpPpp} with
$\pr P'$ which do not affect $l_{1}$ and $l'_{1}$ and $l_{1}\neq i'_{k}\neq l'_{1}$.
$\pr P_{l_{1}l_{2}}$ is an additional transposition between the rows
of the Young tableaux $[0]$ of the shapes $\lambda$ and $\lambda'$
and $n_{ex}=n'_{ex}=k_{m}+1$ , unless $i''_{k}=\lambda_{1}+1$, when
$n_{ex}=n'_{ex}=k_{m}$. Then
\begin{multline*}
\varSigma_{jj'jj'}^{(S-1,S)}(\lambda_{1}-1,\lambda_{1},\lambda_{1}+1,\lambda_{1})=(N-2)!\sum_{k_{m}=0}^{\lambda_{2}}(\lambda_{1}-2)!\lambda_{2}!\binom{\lambda_{1}-2}{k_{m}}\biggl[\binom{\lambda_{2}-1}{k_{m}}\binom{\lambda_{1}}{k_{m}+1}^{-1}\binom{\lambda_{1}-1}{k_{m}+1}^{-1}\\
+\binom{\lambda_{2}-1}{k_{m}-1}\binom{\lambda_{1}}{k_{m}}^{-1}\binom{\lambda_{1}-1}{k_{m}}^{-1}\biggr]=\frac{N!(N-2)!}{f_{S}\lambda_{1}(\lambda_{1}-1)(\lambda_{1}-\lambda_{2}+2)}.
\end{multline*}
If $l_{1}=\lambda_{1}-1$, $l'_{1}=l_{2}=\lambda_{1}$, $l'_{2}=\lambda_{1}+1$,
the Kronecker symbols are satisfied by the permutations \refneq{Pl1l2Pl1pl2pTranPpPpp}
with the same restrictions to $\pr P'$ and $i'_{k}$ as in the previous
case. Since $\pr P_{\lambda_{1}\lambda_{1}-1}\pr P_{\lambda_{1}\lambda_{1}+1}=\pr P_{\lambda_{1}\lambda_{1}+1}\pr P_{\lambda_{1}-1\lambda_{1}+1}$
$n_{ex}=n'_{ex}=k_{m}+1$ , unless $i''_{k}=\lambda_{1}+1$, when
$n_{ex}=n'_{ex}=k_{m}$, as in the previous case, and 
\[
\varSigma_{jj'jj'}^{(S-1,S)}(\lambda_{1}-1,\lambda_{1},\lambda_{1},\lambda_{1}+1)=\varSigma_{jj'jj'}^{(S-1,S)}(\lambda_{1},\lambda_{1}-1,\lambda_{1},\lambda_{1}+1).
\]

The last relevant case is $S'=S-2$. If $l_{1}=l_{2}=\lambda_{1}-1$,
$l'_{1}=l'_{2}=\lambda_{1}$, the Kronecker symbols in \refeq{SigmaSSpjjpjjp}
select the permutations of the form \refneq{PermTranPpPpp} with $\pr P'$
which do not affect $l_{1}$ and $l'_{1}$ and $l_{1}\neq i'_{k}\neq l'_{1}$.
Then
\[
\varSigma_{jj'jj'}^{(S-2,S)}(\lambda_{1}-1,\lambda_{1},\lambda_{1}-1,\lambda_{1})=(N-2)!\sum_{n_{ex}=0}^{\lambda_{2}}(\lambda_{1}-2)!\lambda_{2}!\binom{\lambda_{1}-2}{n_{ex}}\binom{\lambda_{2}}{n_{ex}}\binom{\lambda_{1}}{n_{ex}}^{-1}\binom{\lambda_{1}-2}{n_{ex}}^{-1}=\frac{N!(N-2)!}{f_{S}\lambda_{1}(\lambda_{1}-1)}.
\]
\end{widetext}

If $l_{1}=l'_{2}=\lambda_{1}-1$, $l'_{1}=l_{2}=\lambda_{1}$,the
Kronecker symbols are satisfied by the permutations \refneq{Pl1l2TranPpPpp}
with the same restrictions to $\pr P'$ and $i'_{k}$ as in the previous
case. Now $\pr P_{l_{1}l_{2}}$ is a transposition within the same
row of the Young tableaux $[0]$ of the shapes $\lambda$ or $\lambda'=[\lambda_{1}-2,\lambda_{2}+2]$.
Therefore
\[
\varSigma_{jj'jj'}^{(S-2,S)}(\lambda_{1}-1,\lambda_{1},\lambda_{1},\lambda_{1}-1)=\varSigma_{jj'jj'}^{(S-2,S)}(\lambda_{1},\lambda_{1}-1,\lambda_{1},\lambda_{1}-1)
\]

\section{Calculation of the sums \refneq{SigmaS4}\label{AppS4}}

The sum \refneq{SigmaS4} is expressed using the relations \refneq{RepProd}
and \refneq{InvOrthMat} as

\[
\varSigma_{j_{1}j'_{1}j_{2}j'_{2}}^{(S)}(l,l')=\sum_{\pr P}D_{[0][0]}^{[\lambda]}(\pr P\pr P_{j_{2}j'_{2}}\pr P^{-1})\delta_{l,\pr Pj_{1}}\delta_{l',\pr Pj_{1}'}
\]
Only the sums with $j_{1}\neq j'_{1}$ and $j_{2}\neq j'_{2}$ are
used here. This implies $l\neq l'$.

If $j_{1}=j_{2}$ and $j'_{1}=j'_{2}$, or $j_{1}=j'_{2}$ and $j_{1}=j'_{2}$,
there are $(N-2)!$ permutations $\pr P$ which satisfy the Kronecker
symbols, and the identity \refneq{TransfTransp} leads to
\[
\varSigma_{jj'jj'}^{(S)}(l,l')=\varSigma_{jj'j'j}^{(S)}(l,l')=(N-2)!D_{[0][0]}^{[\lambda]}(\pr P_{ll'}).
\]
Therefore
\[
\varSigma_{jj'jj'}^{(S)}(\lambda_{1}-1,\lambda_{1})=\varSigma_{jj'jj'}^{(S)}(\lambda_{1}+1,\lambda_{1}+2)=(N-2)!,
\]
since $l$ and $l'$ are in the same row of the Young tableau $[0]$
of the shape $\lambda$ {[}see Eq. (I.8){]}. Then 
\begin{equation}
\varSigma_{2}^{(S)}(\lambda_{1}-1,\lambda_{1})=\varSigma_{2}^{(S)}(\lambda_{1}+1,\lambda_{1}+2)=2(N-2)!
\end{equation}
 in \refeq{SumSigVV}.

If $j_{1}=j_{2}$ , but $j_{1}\neq j'_{2}\neq j'_{1}$, 
\begin{multline*}
\varSigma_{jj'_{1}jj'_{2}}^{(S)}(l,l')=\sum_{l\neq l''\neq l'}D_{[0][0]}^{[\lambda]}(\pr P_{ll''})\sum_{\pr P}\delta_{l,\pr Pj}\delta_{l',\pr Pj_{1}'}\delta_{l'',\pr Pj_{2}'}\\
=(N-3)!\sum_{l\neq l''\neq l'}D_{[0][0]}^{[\lambda]}(\pr P_{ll''}).
\end{multline*}
Similarly,
\[
\varSigma_{jj'_{1}j_{2}j}^{(S)}(l,l')=\varSigma_{j_{1}jj_{2}j}^{(S)}(l,l')=\varSigma_{j_{1}jjj'_{2}}^{(S)}(l,l')=\varSigma_{jj'_{1}jj'_{2}}^{(S)}(l,l').
\]

If $l=\lambda_{1}-1$, $l'=\lambda_{1}$, for $l''\leq\lambda_{1}-2$,
$l$ and $l''$ are in the same row of the Young tableau $[0]$ of
the shape $\lambda$, and $D_{[0][0]}^{[\lambda]}(\pr P_{ll''})=1$.
For $l''>\lambda_{1}$, $l$ and $l''$ are in different rows of this
Young tableau, and $D_{[0][0]}^{[\lambda]}(\pr P_{ll''})=-1/\lambda_{1}$
{[}see \refeq{D00Godd}{]}. Then
\[
\varSigma_{jj'_{1}jj'_{2}}^{(S)}(\lambda_{1}-1,\lambda_{1})=(N-3)!\left(\lambda_{1}-2-\frac{\lambda_{2}}{\lambda_{1}}\right).
\]
Similarly,
\[
\varSigma_{jj'_{1}jj'_{2}}^{(S)}(\lambda_{1}+1,\lambda_{1}+2)=(N-3)!\left(\lambda_{2}-3\right).
\]
Then 
\begin{equation}
\begin{aligned}\varSigma_{3}^{(S)}(\lambda_{1}-1,\lambda_{1})= & 4(N-3)!\left(\lambda_{1}-2-\lambda_{2}/\lambda_{1}\right)\\
\varSigma_{3}^{(S)}(\lambda_{1}+1,\lambda_{1}+2)= & 4(N-3)!\left(\lambda_{2}-3\right)
\end{aligned}
\end{equation}
in \refeq{SumSigVV}.

If $j_{1}\neq j_{2}\neq j'_{1}$ and $j_{1}\neq j'_{2}\neq j'_{1}$,
we have\begin{widetext}
\[
\varSigma_{j_{1}j'_{1}j_{2}j'_{2}}^{(S)}(l,l')=\sum_{l\neq l''\neq l'}\sum_{l\neq l'''\neq l',l'''\neq l''}D_{[0][0]}^{[\lambda]}(\pr P_{l''l'''})\sum_{\pr P}\delta_{l,\pr Pj_{1}}\delta_{l',\pr Pj_{1}'}\delta_{l'',\pr Pj_{2}}\delta_{l''',\pr Pj_{2}'}=(N-4)!\sum_{l\neq l''\neq l'}\sum_{l\neq l'''\neq l',l'''\neq l''}D_{[0][0]}^{[\lambda]}(\pr P_{l''l'''}).
\]
Using the same values of the Young matrix elements as in the previous
case, one gets
\begin{equation}
\begin{aligned}\varSigma_{4}^{(S)}(\lambda_{1}-1,\lambda_{1})= & \varSigma_{j_{1}j'_{1}j_{2}j'_{2}}^{(S)}(\lambda_{1}-1,\lambda_{1})=(N-4)!\left[(\lambda_{1}-2)(\lambda_{1}-3)+\lambda_{2}(\lambda_{2}-1)-2(\lambda_{1}-2)\frac{\lambda_{2}}{\lambda_{1}}\right]\\
\varSigma_{4}^{(S)}(\lambda_{1}+1,\lambda_{1}+2)= & \varSigma_{j_{1}j'_{1}j_{2}j'_{2}}^{(S)}(\lambda_{1}+1,\lambda_{1}+2)=(N-4)!\left[(\lambda_{2}-2)(\lambda_{2}-5)+\lambda_{1}(\lambda_{1}-1)\right].
\end{aligned}
\end{equation}
\end{widetext}

\section{Calculation of the sums \refneq{SigmaSSpjj}\label{AppSSpjj}}

The sum \refneq{SigmaSSpjj}, expressed as 
\begin{equation}
\varSigma_{jj}^{(S',S)}(l,l')=(N-1)!\sum_{\pr R}D_{[0][0]}^{[\lambda']}(\pr R)D_{[0][0]}^{[\lambda]}(\pr R)\delta_{l',\pr Rl},\label{SigSpSllp}
\end{equation}
is denoted for consistency with the sum $\varSigma_{jj}^{(S',S)}$
{[}see Eq. (I.A.1){]}, which is its special case for $l=l'=\lambda_{1}$.
Since $D_{[0][0]}^{[\lambda]}(\pr R^{-1})=D_{[0][0]}^{[\lambda]}(\pr R)$
and $\delta_{l',\pr Rl}=\delta_{l,\pr R^{-1}l'}$, the substitution
$\pr R=\pr R^{-1}$ transposes arguments $l$ and $l'$. Therefore
\begin{equation}
\varSigma_{jj}^{(S',S)}(l,l')=\varSigma_{jj}^{(S',S)}(l',l).\label{SymPerllp}
\end{equation}

Consider at first the case $S=S'$. If $l=l'\leq\lambda_{1}$, the
Kronecker symbol in \refeq{SigSpSllp} selects the permutations of
the form \refneq{PermTranPpPpp} with $\pr P'$ which do not affect
$l$ and $i'_{k}\neq l$. Therefore there are $(\lambda_{1}-1)!$
permutations $\pr P'$, $\lambda_{2}!$ permutations $\pr P''$, and
number of distinct choices of the sets of $i'_{k}$ and $i''_{k}$
are given by the binomial coefficients $\binom{\lambda_{1}-1}{n_{ex}}$
and $\binom{\lambda_{2}}{n_{ex}}$, respectively. Then this sum, being
independent of the particular values of $l$, is equal to $\varSigma_{jj}^{(S,S)}(\lambda_{1},\lambda_{1})\equiv\varSigma_{jj}^{(S,S)}$
, calculated in Appendix in Ref. \cite{yurovsky2015},
\[
\varSigma_{jj}^{(S,S)}(l,l)=\frac{N!(N-1)!}{f_{S}\lambda_{1}^{2}}\left[\lambda_{1}-\frac{\lambda_{2}}{\lambda_{1}-\lambda_{2}+2}\right].
\]

If $l\neq l'$, but $l\leq\lambda_{1}$ and $l'\leq\lambda_{1}$,
the Kronecker symbol in \refeq{SigSpSllp} selects permutations 
\begin{equation}
\pr R=\pr P_{ll'}\prod_{k=1}^{k_{m}}\pr P_{i'_{k}i''_{k}}\pr P'\pr P'',\label{PllpTranPpPpp}
\end{equation}
with the same restrictions to $\pr P'$ and $i'_{k}$. Since both
$l$ and $l'$ are in the first row of the Young tableau $[0]$, $\varSigma_{jj}^{(S,S)}(l,l')=\varSigma_{jj}^{(S,S)}(l,l)$
in this case.

If $l=l'>\lambda_{1}$, the Kronecker symbol in \refeq{SigSpSllp}
selects the permutations \refneq{PermTranPpPpp} with $\pr P''$ which
do not affect $l'$ and $i''_{k}\neq l'$. There are $\lambda_{1}!$
permutations $\pr P'$ , $(\lambda_{2}-1)!$ permutations $\pr P''$,
and number of distinct choices of the sets of $i'_{k}$ and $i''_{k}$
are given by the binomial coefficients $\binom{\lambda_{1}}{n_{ex}}$
and $\binom{\lambda_{2}-1}{n_{ex}}$ , respectively. Then \refeq{SigmaSSpjj}
can be represented as,\begin{widetext}
\begin{multline}
\varSigma_{jj}^{(S,S)}(l,l)=(N-1)!\sum_{n_{ex}=0}^{\lambda_{2}}\lambda_{1}!(\lambda_{2}-1)!\binom{\lambda_{1}}{n_{ex}}\binom{\lambda_{2}-1}{n_{ex}}\binom{\lambda_{1}}{n_{ex}}^{-2}=(N-1)!((\lambda_{2}-1)!)^{2}\sum_{n_{ex}=0}^{\lambda_{2}-1}\frac{(\lambda_{1}-n_{ex})!}{(\lambda_{2}-n_{ex}-1)!}\\
=\frac{N!(N-1)!}{f_{S}\lambda_{2}}\left[1-\frac{1}{\lambda_{1}-\lambda_{2}+2}\right].\label{SigSSllpgtL1}
\end{multline}

If $l\neq l'$, but $l>\lambda_{1}$ and $l'>\lambda_{1}$, the Kronecker
symbol in \refeq{SigSpSllp} selects permutations \refneq{PllpTranPpPpp}
with the same restrictions to $\pr P''$ and $i''_{k}$ as in the
case of $l=l'>\lambda_{1}$. Since both $l$ and $l'$ are in the
second row of the Young tableau $[0]$, $\varSigma_{jj}^{(S,S)}(l,l')=\varSigma_{jj}^{(S,S)}(l,l)$
in this case.

If $l\leq\lambda_{1}<l'$, the permutations \refneq{PllpTranPpPpp}
satisfy the Kronecker symbol in \refeq{SigSpSllp} if $\pr P'$ do
not affect $l$ and $i'{}_{k}\neq l$. If $i''_{k}\neq l'$ for any
$k$, $\pr P_{ll'}$ is an additional transposition between the rows
of the Young tableau $[0]$, and $n_{ex}=k_{m}+1$. Otherwise, if
$i''_{k}=l'$, since $\pr P_{ll'}\pr P_{l'i'_{k}}=\pr P_{li'_{k}}\pr P_{ll'}$,
$n_{ex}=k_{m}$. Then
\[
\varSigma_{jj}^{(S,S)}(l,l')=(N-1)!\sum_{k_{m}=0}^{\lambda_{2}}(\lambda_{1}-1)!\lambda_{2}!\binom{\lambda_{1}-1}{k_{m}}\left[\binom{\lambda_{2}-1}{k_{m}}\binom{\lambda_{1}}{k_{m}+1}^{-2}+\binom{\lambda_{2}-1}{k_{m}-1}\binom{\lambda_{1}}{k_{m}}^{-2}\right]=\frac{N!(N-1)!}{f_{S}\lambda_{1}(\lambda_{1}-\lambda_{2}+2)}.
\]

The next case is $S'=S-1$. If $l=l'=\lambda_{1}$, the sum was calculated
in Appendix in Ref. \cite{yurovsky2015},
\[
\varSigma_{jj}^{(S-1,S)}(\lambda_{1},\lambda_{1})\equiv\varSigma_{jj}^{(S-1,S)}=\frac{N!(N-1)!}{f_{S}\lambda_{1}}.
\]

If $l=l'=\lambda_{1}-1$, the Kronecker symbol in \refeq{SigSpSllp}
selects permutations 

\begin{equation}
\pr R=\prod_{k=1}^{k_{m}}\pr P_{i'_{k}i''_{k}}\pr P_{\lambda_{1}i_{0}}\pr P'\pr P'',\label{PermTrani0PpPpp}
\end{equation}
if $\pr P'$ are permutations of the $\lambda_{1}-2$ first symbols,
$\pr P''$ are permutations of the $\lambda_{2}$ last symbols, $i_{0}\neq\lambda_{1}-1$,
$i'_{k}\leq\lambda_{1}-2$, and $i''_{k}>\lambda_{1}$. The numbers
of transpositions $n_{ex}$ and $n'_{ex}$ between rows of the Young
tableaux $[0]$ of the shapes $\lambda$ and $\lambda'$, respectively,
depend on $i_{0}$. If $i_{0}=\lambda_{1}$, $n_{ex}(i_{0})=n'_{ex}(i_{0})=k_{m}$
. If $i_{0}\leq\lambda_{1}-2$, $n_{ex}(i_{0})=k_{m}$ , $n'_{ex}(i_{0})=k_{m}+1$,
unless $i_{0}=i'_{k}$ for any $k$. In the last case, $n_{ex}(i_{0})=k_{m}$
and, since $\pr P_{i'_{k}i''_{k}}\pr P_{\lambda_{1}i'_{k}}=\pr P_{\lambda_{1}i'_{k}}\pr P_{\lambda_{1}i''_{k}}$,
$n'_{ex}(i_{0})=k_{m}$. Similarly, if $i_{0}>\lambda_{1}$, $n'_{ex}(i_{0})=k_{m}$
, $n_{ex}(i_{0})=k_{m}+1$, unless $i_{0}=i''_{k}$, when $n_{ex}(i_{0})=k_{m}$.
Thus for $2k_{m}+1$ values of $i_{0}$ we have $n_{ex}(i_{0})=n'_{ex}(i_{0})=k_{m}$,
for $\lambda_{1}-2-k_{m}$ values $n_{ex}(i_{0})=k_{m}$, $n'_{ex}(i_{0})=k_{m}+1$,
and for $\lambda_{2}-k_{m}$ values $n_{ex}(i_{0})=k_{m}+1$, $n'_{ex}(i_{0})=k_{m}$.
Then the sum \refneq{SigmaSSpjj} is expressed as
\begin{multline*}
\varSigma_{jj}^{(S-1,S)}(\lambda_{1}-1,\lambda_{1}-1)=(N-1)!\sum_{k_{m}=0}^{\lambda_{2}}(\lambda_{1}-2)!\lambda_{2}!\binom{\lambda_{1}-2}{k_{m}}\binom{\lambda_{2}}{k_{m}}\biggl[(2k_{m}+1)\binom{\lambda_{1}}{k_{m}}^{-1}\binom{\lambda_{1}-1}{k_{m}}^{-1}\\
-(\lambda_{1}-2-k_{m})\binom{\lambda_{1}}{k_{m}}^{-1}\binom{\lambda_{1}-1}{k_{m}+1}^{-1}-(\lambda_{2}-k_{m})\binom{\lambda_{1}}{k_{m}+1}^{-1}\binom{\lambda_{1}-1}{k_{m}}^{-1}\biggl]=\frac{N!(N-1)!}{f_{S}\lambda_{1}(\lambda_{1}-1)^{2}}.
\end{multline*}

If $l=\lambda_{1}$, $l'=\lambda_{1}-1$, the Kronecker symbol in
\refeq{SigSpSllp} selects permutations \refneq{PllpTranPpPpp} if
$\pr P'$ are permutations of the $\lambda_{1}-1$ first symbols,
$\pr P''$ are permutations of the $\lambda_{2}$ last symbols, $i'_{k}\leq\lambda_{1}-1$,
and $i''_{k}>\lambda_{1}$ . Now $n_{ex}=k_{m}$ , $n'_{ex}=k_{m}+1$
unless $i'_{k}=\lambda_{1}-1$ for any $k$, when $n_{ex}=n'_{ex}=k_{m}$.
Then
\begin{multline*}
\varSigma_{jj}^{(S-1,S)}(\lambda_{1},\lambda_{1}-1)=(N-1)!\sum_{k_{m}=0}^{\lambda_{2}}(\lambda_{1}-1)!\lambda_{2}!\binom{\lambda_{2}}{k_{m}}\binom{\lambda_{1}}{k_{m}}^{-1}\left[\binom{\lambda_{1}-2}{k_{m}-1}\binom{\lambda_{1}-1}{k_{m}}^{-1}-\binom{\lambda_{1}-2}{k_{m}}\binom{\lambda_{1}-1}{k_{m}+1}^{-1}\right]\\
=-\frac{N!(N-1)!}{f_{S}\lambda_{1}(\lambda_{1}-1)}.
\end{multline*}
\end{widetext}

A general relation can be derived for $l\geq\lambda_{1}+1$ and arbitrary
$l'$, when the Kronecker symbol in \refeq{SigSpSllp} is satisfied
by 
\begin{equation}
\pr R=\pr P_{l'N}\pr P\pr P_{lN}\label{PlpNPPlN}
\end{equation}
 with arbitrary $\pr P\in\pr S_{N-1}$, such that $\pr PN=N$. Then
$D_{[0][0]}^{[\lambda]}(\pr R)=D_{[0][0]}^{[\lambda]}(\pr P_{l'N}\pr P)$,
since both $l$ and $N$ are in the second row of the Young tableau
$[0]$ of the shape $\lambda$ {[}see Eq. (I.8){]}. If $S'<S$ and
$\lambda'_{1}<\lambda_{1}$, $D_{[0][0]}^{[\lambda']}(\pr R)=D_{[0][0]}^{[\lambda']}(\pr P_{l'N}\pr P)$
by the same reason. Using \refeq{RepProd}, the sum can be expressed
as
\begin{multline*}
\varSigma_{jj}^{(S',S)}(l,l')=(N-1)!\sum_{r,r'}D_{[0]r}^{[\lambda]}(\pr P_{l'N})D_{[0]r'}^{[\lambda']}(\pr P_{l'N})\\
\times\sum_{\pr P\in\pr S_{N-1}}D_{r[0]}^{[\lambda]}(\pr P)D_{r'[0]}^{[\lambda']}(\pr P).
\end{multline*}
As $\pr P$ are elements of the subgroup $\pr S_{N-1}$ of permutations
of $N-1$ first symbols, a reduction to subgroup (see \cite{kaplan})
can be used, $D_{rt}^{[\lambda]}(\pr P)=D_{\bar{r}\bar{t}}^{[\bar{\lambda}]}(\pr P)$,
where the Young tableaux $\bar{r}$ and $\bar{t}$, corresponding
to the same Young diagram, $\bar{\lambda}$, are obtained by the removal
of the symbol $N$ from the tableaux $r$ and $t$, respectively.
($D_{rt}^{[\lambda]}(\pr P)=0$ if $\bar{r}$ and $\bar{t}$ correspond
to different Young diagrams due to different placement of the symbol
$N$ in $r$ and $t$ .) The summation over $\pr P$ can be then performed
using the orthogonality relation (see \cite{kaplan,pauncz_symmetric})
\begin{equation}
\sum_{\pr Q\in\pr S_{N-1}}D_{\bar{t}'\bar{r}'}^{[\bar{\lambda}']}(\pr Q)D_{\bar{t}\bar{r}}^{[\bar{\lambda}]}(\pr Q)=\frac{N!}{f_{\bar{\lambda}_{1}-\bar{\lambda}_{2}}(N-1)}\delta_{\bar{t}\bar{t}'}\delta_{\bar{r}\bar{r}'}\delta_{\bar{\lambda}\bar{\lambda}'},\label{OrthNm1}
\end{equation}
where $f_{S}(N-1)$ is the representation dimension for $N-1$ particles.
The symbol $N$ is placed in the end of the second row in the Young
tableau $[0]$. Therefore, $\bar{\lambda}=[\lambda_{1},\lambda_{2}-1]$,
$\bar{\lambda}'=[\lambda'_{1},\lambda'_{2}-1]$, and $\bar{\lambda}=\bar{\lambda}'$
only if $\lambda_{1}=\lambda'_{1}$, or $S=S'$. As a result,
\[
\varSigma_{jj}^{(S',S)}(l,l')\propto\delta_{\bar{\lambda}\bar{\lambda}'}=0
\]
whenever $l\geq\lambda_{1}+1$ and $S'<S$. Due to \refeq{SymPerllp},
the sum vanish whenever $l'\geq\lambda_{1}+1$ for arbitrary $l$
too. This general relation provides the sums appearing in the present
calculations
\begin{multline*}
\varSigma_{jj}^{(S-1,S)}(\lambda_{1}-1,\lambda_{1}+1)=\varSigma_{jj}^{(S-1,S)}(\lambda_{1},\lambda_{1}+1)\\
=\varSigma_{jj}^{(S-1,S)}(\lambda_{1}+1,\lambda_{1})=\varSigma_{jj}^{(S-1,S)}(\lambda_{1}+1,\lambda_{1}+1)=0.
\end{multline*}

Another general relation restricts difference between $S$ and $S'$.
Equations \refneq{RepProd}, \refneq{PlpNPPlN} , and the reduction
to subgroup lead to $D_{[0][0]}^{[\lambda]}(\pr R)=\sum_{r,t}D_{[0]r}^{[\lambda]}(\pr P_{l'N})D_{\bar{r}\bar{t}}^{[\bar{\lambda}]}(\pr P)D_{t[0]}^{[\lambda]}(\pr P_{lN})$
and $D_{[0][0]}^{[\lambda']}(\pr R)=\sum_{r',t'}D_{[0]r'}^{[\lambda']}(\pr P_{l'N})D_{\bar{r}'\bar{t}'}^{[\bar{\lambda}']}(\pr P)D_{t'[0]}^{[\lambda']}(\pr P_{lN})$.
Then the sum \refneq{SigSpSllp} contains 
\[
\sum_{\pr P\in\pr S_{N-1}}D_{\bar{t}'\bar{r}'}^{[\bar{\lambda}']}(\pr P)D_{\bar{t}\bar{r}}^{[\bar{\lambda}]}(\pr P)\propto\delta_{\bar{\lambda}\bar{\lambda}'}
\]
{[}see \refeq{OrthNm1}{]}. Now the symbol $N$ can be placed in the
end of either row of the Young tableaux. Then both $\bar{\lambda}=[\lambda_{1},\lambda_{2}-1]$
and $\bar{\lambda}=[\lambda_{1}-1,\lambda_{2}]$ are allowed, as well
as two similar $\bar{\lambda}'$. Therefore, $|\lambda_{1}-\lambda'_{1}|\leq1$,
or $|S-S'|\leq1$. This provides the sums appearing in the present
calculations
\begin{multline*}
\varSigma_{jj}^{(S-2,S)}(\lambda_{1}-1,\lambda_{1}-1)=\varSigma_{jj}^{(S-2,S)}(\lambda_{1}-1,\lambda_{1})\\
=\varSigma_{jj}^{(S-2,S)}(\lambda_{1},\lambda_{1}-1)=\varSigma_{jj}^{(S-2,S)}(\lambda_{1},\lambda_{1})=0.
\end{multline*}

\begin{acknowledgments}
The author gratefully acknowledges useful conversations with N. Davidson,
V. Fleurov, I. G. Kaplan, and E. Sela. 
\end{acknowledgments}

\end{document}